\documentclass[%
 reprint,
nofootinbib,
 amsmath,amssymb,
 aps
]{revtex4-1}
\usepackage{multirow}
\usepackage{graphicx}
\usepackage{dcolumn}
\usepackage{bm}
\usepackage{amsmath}
\usepackage{amssymb}
\usepackage{mathrsfs}
\usepackage{amsfonts}
\usepackage{color, soul}

\def\oversim#1#2{\lower0.5ex\vbox{\baselineskip0pt\lineskip0pt
                 \lineskiplimit0pt\everycr{}\tabskip0pt
                 \halign{$\mathsurround0pt #1\hfil##\hfil$\crcr #2\crcr\sim\crcr}}}

\begin{document}

\preprint{ARXIV/1612.08522}

\title{The diphoton signal of the light Higgs boson in Natural NMSSM}

\author{Junjie Cao$^{1,2}$, Xiaofei Guo$^1$, Yangle He$^1$, Peiwen Wu$^3$, Yang Zhang$^4$}

\affiliation{ $^1$  College of Physics and Materials Science,
        Henan Normal University, Xinxiang 453007, China \\
 $^2$ Department of Applied Physics, Xi'an Jiaotong University, Xi'an 710049, China \\
  $^3$ School of Physics, KIAS, 85 Hoegiro, Seoul 02455, Republic of Korea \\
  $^4$ State Key Laboratory of Theoretical Physics,
      Institute of Theoretical Physics, Academia Sinica, Beijing 100190,
      China }

\email{junjiec@itp.ac.cn, \,peiwen.wu123@gmail.com}


\begin{abstract}
Natural Next-to-Minimal Supersymmetric Standard Model (nNMSSM) is featured by predicting one CP-even Higgs boson
satisfying $m_{h_1} \lesssim 120 \,{\rm GeV}$ and Higgsinos lighter than about 300 GeV, and consequently
the cross section for DM-nucleon scattering in this scenario is usually quite large.
We study the diphoton signal of the light Higgs boson in nNMSSM by considering the tight constraints
from the latest LUX and PandaX-II experiments, and we conclude that the optimal value of the signal rate at 8 TeV LHC is greatly
reduced in comparison with earlier predictions. For example, previous studies indicated that
the rate may exceed $120 \,{\rm fb}$ for $m_{h_1} \simeq 80 \,{\rm GeV}$, while it is at most
$25 \,{\rm fb}$  if the lightest neutralino in the scenario is fully responsible for the measured DM relic density.
We also investigate the case of $m_{h_1} \simeq 98 \,{\rm GeV}$ which is hinted by the excesses of the LEP analysis on
$Z \bar{b} b$ signal and the CMS analysis on the diphoton signal. We conclude that nNMSSM can explain simultaneously the excesses
at $1\sigma$ level without violating any known constraints.
\end{abstract}

\pacs{}

\maketitle

\section{\label{intro}Introduction}

The hierarchy problem of the Standard Model (SM) usually implies a more complex structure in Higgs sector to explain the electroweak symmetry breaking. In the present context
of continuing efforts paid to search for new particle at the upgraded LHC, it is one of the priority to look for extra Higgs bosons.
Since experimentally photon is a very clean object and can be reconstructed with a very high precision, the diphoton signal of the bosons has been considered
as the golden channel in the search, especially when the bosons are moderately light, and one remarkable achievement in this direction is the great discovery of
a  $125 \,{\rm GeV}$ Higgs boson in 2012 \cite{ATLAS-Higgs,CMS-Higgs}.
On the theoretical side, supersymmetric theory provides an elegant way to stabilize the electroweak scale and is regarded as one of the most promising
candidates of new physics.  However, in order to accommodate a Higgs boson with its mass around $125 \,{\rm GeV}$ in the Minimal Supersymmetric Standard Model (MSSM),
the contribution to the mass from higher order correction is required to be very close to its tree level value which seems rather unnatural \cite{MSSM-Higgs}.
As a result, non-minimal supersymmetric theories have drawn a lot attention in recent years. In this work we concentrate on the Next-to-Minimal
Supersymmetric Standard Model (NMSSM) \cite{NMSSM-Report} which is the simplest extension of the MSSM with one singlet Higgs field. To be more specific, we study the diphoton
signal of the lightest CP-even Higgs boson in the most attractive scenario of the NMSSM, which is dubbed as natural NMSSM (nNMSSM) \cite{King-nNMSSM}.

The essential feature of nNMSSM is that among the three CP-even Higgs bosons predicted
by the NMSSM, the next to lightest one corresponds to the $125 \,{\rm GeV}$ Higgs, which is usually called the SM-like Higgs boson,
and in order to achieve this the Higgsino mass parameter $\mu$ is preferred to be lighter than about
$300 \,{\rm GeV}$ \cite{King-nNMSSM,NMSSM-125-1}. In this scenario, the Higgs mass can be lifted by both singlet-doublet-doublet Higgs coupling and singlet-doublet Higgs mixing,
and consequently its value can be easily enhanced to $125 \,{\rm GeV}$ without the large radiative correction
\cite{NMSSM-125-1,NMSSM-125-2,NMSSM-125-3,NMSSM-125-4,NMSSM-125-5}. This fact along with the condition
$\mu \lesssim 300 \,{\rm GeV}$ makes the theory rather natural in predicting $Z$ boson mass \cite{nNMSSM-LHC}.

It should be noted that the lightest CP-even Higgs in nNMSSM (denoted by $h_1$ hereafter) is rather peculiar. First, since it is
lighter than about $120 \,{\rm GeV}$, its properties have been tightly limited by the LEP experiments \cite{Light-Higgs-Constraint}
and also by some analyses at the LHC \cite{Aad:2014ioa,CMS-HIG-14-037}. Considering that the invariant mass of the diphoton signal can be determined
rather precisely in experiment, its future observation at colliders will provide a robust clue to popular singlet extensions of the SM, and may also be used to distinguish the NMSSM
from the MSSM since the latter can not predict such a spectrum after considering relevant experimental constraints.  Second,
the recent dark matter (DM) direct detection experiments such as LUX and PandaX-II have imposed
strong constraints on supersymmetric models \cite{nNMSSM-LUX}. In this case, the existence of a light CP-even Higgs boson is favored to relax the constraints
since in certain parameter space, its contribution to the spin-independent (SI) cross section of DM-nucelon scattering is comparable but with opposite sign to that of the SM-like Higgs boson
so that the cross section is greatly reduced \cite{nNMSSM-LUX}. Finally,
there are experimental hints on the existence of a light scalar. For example, both the LEP analysis on
$Z \bar{b} b$ signal \cite{Barate:2003sz} and the recent CMS analysis on diphoton signal at LHC Run-I \cite{CMS-HIG-14-037} have observed a $2\sigma$ excess
over the corresponding background, which may be explained simultaneously by the presence of a CP-even Higgs boson with mass around
$98 \, {\rm GeV}$ (about NMSSM explanation of the $Z b \bar{b}$ excess, see \cite{98GeV-Higgs-1,98GeV-Higgs-2,98GeV-Higgs-3,98GeV-Higgs-4,
98GeV-Higgs-5,98GeV-Higgs-6,98GeV-Higgs-7,98GeV-Higgs-8,98GeV-Higgs-9}).

So far there are numerous discussions on the properties of $h_1$ in nNMSSM and its future detection at the LHC \cite{Light-Higgs-1,
Light-Higgs-2,Light-Higgs-3,Light-Higgs-4,Light-Higgs-5,Light-Higgs-6,Light-Higgs-7,Light-Higgs-8,Light-Higgs-9,Light-Higgs-10,
Light-Higgs-11,Light-Higgs-12,Light-Higgs-13,Light-Higgs-14,Light-Higgs-15,Light-Higgs-16,Light-Higgs-17,Light-Higgs-18,Diphoton-1,
Diphoton-2,Diphoton-3,Diphoton-4,Diphoton-5,Diphoton-6,Diphoton-7,Diphoton-8,Diphoton-9,Diphoton-10,Diphoton-11,Diphoton-12,Diphoton-13,
Diphoton-14}, especially the diphoton signal of $h_1$ at the LHC was intensively studied in \cite{Diphoton-1,
Diphoton-2,Diphoton-3,Diphoton-4,Diphoton-5,Diphoton-6,Diphoton-7,Diphoton-8,Diphoton-9,Diphoton-10,Diphoton-11,Diphoton-12,Diphoton-13,
Diphoton-14,SchmidtHoberg:2012yy,Barbieri:2013nka}. These studies indicated that there exist some parameter regions where the couplings of $h_1$ with down-type quarks
are more suppressed than those with up-type quarks and vector bosons so that the branching ratio of
$h_1 \to \gamma \gamma$ can be greatly enhanced. In this case,
the diphoton rate may be several times larger than its SM prediction for the same scalar mass \cite{Diphoton-2}.  In this work, we update previous
studies in this subject by considering the constraints from DM physics, especially the impacts of the recent LUX
and PandaX-II experiments \cite{Akerib:2016vxi,Fu:2016ega,Tan:2016zwf} on the theory. Our results indicate that the DM experiments
are very efficient in excluding the parameter space of nNMSSM even if we assume that the lightest
neutralino in the scenario constitutes only a small fraction of the DM in the Universe. As a result, previous results on the diphoton signal are exorbitantly
optimistic. For example, compared with the latest study on the diphoton rate in \cite{Diphoton-14}, we find that  the maximal
theoretical prediction of the rate for $m_{h_1} = 80 \, {\rm GeV}$ drops from more than $120 \, {\rm fb}$ to about $25 \, {\rm fb}$
after including the constraints. We also consider the case of $m_{h_1} \simeq 98 \, {\rm GeV}$ to study whether
nNMSSM can explain simultaneously the excesses reported by the LEP and CMS experiments. We conclude that even if the
lightest neutralino is required to be solely responsible for the observed DM relic density,
nNMSSM can still explain the excesses at $1\sigma$ level without violating any known constraint.

This paper is organized as follows. In Section 2, we recapitulate the basics of the NMSSM
which are helpful to understand the results of this work. In Section 3 we investigate the diphoton rate of $h_1$ by
performing an intensive scan over the vast parameter space of the NMSSM with various constraints. Different features
of the rate are shown by deliberate figures. In Section 4 we turn to investigate wether nNMSSM can explain simultaneously the
excesses observed by LEP and CMS experiments.  Finally, we draw our conclusions in Section 5.

\section{Basics of the NMSSM }

As one of the most economical extensions of the MSSM,
the NMSSM introduces one gauge singlet Higgs superfield in its matter content and
usually adopts a $Z_3$ symmetry in the construction of its
superpotential to avoid the appearance of dimensional parameters.
In this work we impose the $Z_3$ symmetry and the NMSSM superpotential and soft breaking  terms in Higgs sector are  \cite{NMSSM-Report}
\begin{eqnarray}
  W^{\rm NMSSM} &=& W_F + \lambda\hat{H_u} \cdot \hat{H_d} \hat{S}
  +\frac{1}{3}\kappa \hat{S^3},\\
  V^{\rm NMSSM}_{\rm soft} &=& \tilde m_u^2|H_u|^2 + \tilde m_d^2|H_d|^2
  +\tilde m_s^2|S|^2 \\ \nonumber
 &+& ( \lambda A_{\lambda} SH_u\cdot H_d
  +\frac{1}{3}\kappa A_{\kappa} S^3 + h.c.),
\end{eqnarray}
where $W_F$ is the superpotential of the MSSM without the $\mu$-term and $\hat{H_u}$, $\hat{H_d}$
and $\hat{S}$ are Higgs superfields with $H_u$, $H_d$ and $S$ being their scalar components, respectively.
The dimensionless coefficients $\lambda$ and $\kappa$ parameterize the strengthes of the Higgs self couplings,
and the dimensional quantities $\tilde{m}_{u}$, $\tilde{m}_{d}$, $\tilde{m}_{s}$, $A_\lambda$ and $A_\kappa$ are soft-breaking
parameters. In practice,  the squared masses $\tilde{m}_{u}^2$, $\tilde{m}_d^2$ and $\tilde{m}_s^2$ are traded for $m_Z$,
$\tan \beta \equiv v_u/v_d$
and $\mu \equiv \lambda v_s $ as theoretical inputs
after considering the electroweak symmetry breaking conditions \cite{NMSSM-Report}, where $v_u, v_d, v_s$ represent the vacuum expectation value (vev) of $H_u,H_d,S$ fields, respectively.

Due to the presence of the superfield $\hat{S}$, the NMSSM contains one more complex Higgs field $S$ compared to the MSSM, and
a singlino field which is the fermion component of $\hat{S}$.  Consequently in the NMSSM there are three (two) CP-even (CP-odd) Higgs particles
corresponding to the mixings of the real (imaginary) parts of the $H_u, H_d, S$ fields, and five neutralinos composed of bino,
wino, higgsino and singlino fields. Throughout this paper we denote these particles by $h_i$ ($i=1,2,3$), $A_i$ ($i=1,2$) and $\tilde{\chi}_i^0$ ($i=1,\cdots 5$) respectively with the convention $m_{h_1} < m_{h_2} < m_{h_3}$, $m_{A_1} < m_{A_2}$ and $m_{\tilde{\chi}_1^0} < m_{\tilde{\chi}_2^0} < \cdots <  m_{\tilde{\chi}_5^0}$.
In the following we briefly introduce the key features of these particles, which is helpful to understand the results of this work.

\subsection{The Higgs sector}

In order to present the mass matrices of the Higgs fields in a physical way, we rotate the fields $H_u$ and $H_d$ as \cite{NMSSM-Report}
\begin{eqnarray}
H_1 &=& \cos\beta H_u + \varepsilon \sin\beta H_d^*,\\ \nonumber
H_2 &=& \sin\beta H_u - \varepsilon \cos\beta H_d^*, ~~ H_3 = S,
\end{eqnarray}
where $\varepsilon$ is an antisymmetric tensor with $\varepsilon_{12}=-\varepsilon_{21}=1$ and $\varepsilon_{11}=\varepsilon_{22}=0$.
After this rotation, the redefined fields $H_i$ ($i=1,2,3$) have the following form
\begin{eqnarray}
H_1 &=& \left ( \begin{array}{c} H^+ \\
       \frac{S_1 + i P_1}{\sqrt{2}}
        \end{array} \right),~~
H_2 = \left ( \begin{array}{c} G^+
            \\ v + \frac{ S_2 + i G^0}{\sqrt{2}}
            \end{array} \right), \\ \nonumber
H_3  &=& v_s +\frac{1}{\sqrt{2}} \left(  S_3 + i P_2 \right).
\label{fields}
\end{eqnarray}
where $H_2$ corresponds to the SM Higgs doublet with $G^+, G^0$ being the Goldstone bosons eaten by $W$ and $Z$ bosons
respectively, and $H_1$ represents a new $SU(2)_L$ doublet scalar field with no
coupling to $W$ and $Z$ bosons at tree-level.

In the CP-conserving NMSSM, the fields $S_1$, $S_2$ and $S_3$ mix to form three physical CP-even Higgs bosons. In the basis ($S_1$, $S_2$, $S_3$), the elements of the
corresponding mass matrix are given by \cite{NMSSM-Report}
\begin{eqnarray}
  M^2_{11} &=&  M^2_A + (m^2_Z -\lambda^2 v^2) \sin^2 2\beta, \nonumber \\
  M^2_{12} &=&  -\frac{1}{2}(m^2_Z-\lambda^2 v^2)\sin4\beta, \nonumber \\
  M^2_{13} &=&  -(\frac{M^2_A}{2\mu/\sin2\beta}+\kappa v_s) \lambda v\cos2\beta, \nonumber \\
  M^2_{22} &=&  m_Z^2\cos^2 2\beta +\lambda^2v^2\sin^2 2\beta, \nonumber \\
  M^2_{23} &=&  2\lambda\mu v[1-(\frac{M_A}{2\mu/\sin2\beta})^2 -\frac{\kappa}{2\lambda}\sin2\beta], \nonumber \\
  M^2_{33} &=&  \frac{1}{4}\lambda^2 v^2(\frac{M_A}{\mu/\sin2\beta})^2 +\kappa v_s A_{\kappa} \\ \nonumber
  &+& 4(\kappa v_s)^2 -\frac{1}{2}\lambda\kappa v^2 \sin 2\beta,
  \label{MS}
\end{eqnarray}
where  $M_A$ represents the mass scale of the doublet field $H_1$, and is given by \begin{eqnarray}
M^2_A \equiv m_{P_1 P_1}^2 = \frac{2\mu}{\sin2\beta}(A_{\lambda}+\kappa v_s).
\end{eqnarray}
This mass matrix indicates that the squared mass of the SM Higgs field $S_2$, $ M^2_{22}$,
gets an additional contribution  $\lambda^2v^2$ in comparison with the MSSM expression, and for $\lambda^2v^2>M_Z^2$ its
tree-level value is maximized with $\tan \beta \simeq 1$. This matrix also indicates
that if the relation $m_{S_3 S_3}^2 < m_{S_2 S_2}^2$ holds, the mixing between the fields $S_2$ and $S_3$ can further
enhance the mass of the SM-like Higgs boson. In this case, $h_1$ is a singlet-dominate scalar while $h_2$ plays
the role of the SM Higgs boson. Benefiting from the above contributions, $m_{h_2} \simeq 125 \, {\rm GeV}$
does not necessarily require a large radiative contribution from stop loops \cite{NMSSM-125-1,NMSSM-125-2,NMSSM-125-3,NMSSM-125-4,NMSSM-125-5}.
Due to this attractive feature, the scenario with $h_2$ corresponding to the SM-like Higgs boson was usually
called natural NMSSM \cite{King-nNMSSM}.

The mass matrix in Eq.(\ref{MS}) can be diagonalized by an orthogonal $3 \times 3$ matrix $V$, and consequently the physical states
$h_i$ are given by
\begin{eqnarray}
h_i=\sum_{j=1}^3 V_{ij} S_j.
\end{eqnarray}
With this notation and also noting the fact that current LHC data have required the properties of the $125 \, {\rm GeV}$ boson to highly mimic
those of the SM Higgs boson, one can infer that the normalized couplings of $h_1$ in nNMSSM with SM particles
take following form
\begin{eqnarray}
C_{h_1 u \bar{u}} &\simeq& V_{11} \cot \beta + V_{12},\\ \nonumber
C_{h_1 d \bar{d}} &\simeq&  V_{11} \tan \beta + V_{12}, \quad C_{h_1 V V} = V_{12}.
\label{h1-couplings}
\end{eqnarray}
Since so far sparticles and charged Higgs bosons are preferred to be heavy by the LHC searches for new particles, their influence on the $h_1$ couplings
is usually negligible \cite{King-nNMSSM}. Therefore we can approximate the diphoton rate of $h_1$ at the LHC by the following formula
\begin{eqnarray}
\sigma_{\gamma \gamma} &\equiv & \sigma (g g \to h_1 \to \gamma \gamma) \nonumber \\
&=& \sigma ( g g \to h_1 ) \times Br(h_1 \to \gamma \gamma) \nonumber \\
&\simeq & C_{h_1 u \bar{u}}^2 \sigma_{\rm SM} ( g g \to h_1 ) \frac{C_{h_1 u \bar{u}}^2  \Gamma^{\rm SM}_{\gamma \gamma}}{\Gamma_{tot}} \nonumber \\
&\simeq &  C_{h_1 u \bar{u}}^4 \frac{\Gamma^{\rm SM}_{tot}}{\Gamma_{tot}} \times  \sigma_{\rm SM} ( g g \to h_1 ) Br_{\rm SM} (h_1 \to \gamma \gamma)  \label{Diphoton-Exp-1}
\end{eqnarray}
where $\sigma_{\rm SM}$ and $Br_{\rm SM}$ are the cross section and branching ratio of a SM Higgs boson with same mass as $h_1$ respectively, and $\Gamma_{tot}$
is the total width of $h_1$ given by
\begin{eqnarray}
\Gamma_{tot} &=&  \Gamma_{b \bar{b}} + \Gamma_{c \bar{c}}
+ \Gamma_{\tau \bar{\tau}} + \Gamma_{g g} + \cdots \nonumber \\
&\simeq & C_{h_1 d \bar{d}}^2  (\Gamma^{\rm SM}_{b \bar{b}} + \Gamma^{\rm SM}_{\tau \bar{\tau}} )
+ C_{h_1 u \bar{u}}^2 ( \Gamma^{\rm SM}_{c \bar{c}} + \Gamma^{\rm SM}_{g g}) \nonumber \\
&+& \cdots.   \label{Diphoton-Exp-2}
\end{eqnarray}
Eq.(\ref{Diphoton-Exp-1}) and Eq.(\ref{Diphoton-Exp-2})  indicate that the diphoton rate of $h_1$ in nNMSSM may be moderately large if
$C_{h_1 b \bar{b}} \simeq 0$ (achieved by accidental cancelation between $V_{11} \tan \beta$ and $V_{12}$) and
meanwhile $C_{h_1 u \bar{u}}$ is not suppressed too much. This is possible in some corners of the NMSSM parameter space \cite{Diphoton-2,Diphoton-14}
which is what we are interested in.  These equations also imply that an enhanced diphoton rate is
usually associated with a suppressed $b \bar{b}$ signal of $h_1$. This correlation can affect our explanation of the $98 \, {\rm GeV}$ excesses observed by
LEP and CMS experiments.  Throughout this work, we use the public code SusHi 1.5 \cite{Harlander:2012pb} to obtain the NNLO gluon fusion production cross section for
a SM-like Higgs boson, and multiply it by the normalized $g g h_1$ coupling given by NMSSMTools \cite{NMSSMTools} to get
$\sigma ( g g \to h_1 )$. We checked that the cross section for the bottom fusion production of $h_1$ is usually significantly smaller than
$\sigma ( g g \to h_1 )$, and thus can be safely neglected.

Similarly the fields $P_1$ and $P_2$ mix to form CP-odd Higgs bosons $A_1$ and $A_2$. One subtle point about the pseudoscalars is that the LHC search
for non-standard Higgs bosons has required the doublet-dominated one to be heavier than about $400 \, {\rm GeV}$, while the dominated one may still be arbitrarily light. An important application of this feature is that the mass of the singlet-dominated
pseudoscalar can be tuned around $2 m_{\tilde{\chi}_1^0}$ so that a moderately light $\tilde{\chi}_1^0$ can annihilate via the resonance
to result in a correct relic density and also sizable cross section for DM annihilation in Galactic Center \cite{Cao:2014efa,Cao:2015loa}.

\subsection{The neutralino sector}

The neutralino sector of the  NMSSM consists of the fields bino $\tilde{B}^0$, wino $\tilde{W}^0$, higgsinos $\tilde{H}_{d,u}^0$ and singlino
$\tilde{S}^0$. Taking the basis $\psi^0 = (-i \tilde{B}^0, - i \tilde{W}^0, \tilde{H}_{d}^0, \tilde{H}_{u}^0,
\tilde{S}^0)$, one has the following symmetric neutralino mass matrix
\begin{equation}
{\cal M} = \left(
\begin{array}{ccccc}
M_1 & 0 & -\frac{g_1 v_d}{\sqrt{2}} & \frac{g_1 v_u}{\sqrt{2}} & 0 \\
  & M_2 & \frac{g_2 v_d}{\sqrt{2}} & - \frac{g_2 v_u}{\sqrt{2}} &0 \\
& & 0 & -\mu & -\lambda v_u \\
& & & 0 & -\lambda v_d\\
& & & & \frac{2 \kappa}{\lambda} \mu
\end{array}
\right), \label{eq:MN}
\end{equation}
where $M_1$ and $M_2$ are bino and wino soft breaking mass respectively. With the rotation matrix $N$ for the mass matrix,
neutralino mass eigenstates are given by
\begin{eqnarray}
\tilde{\chi}_i^0 = \sum_{j=1}^5 N_{ij}\psi_j^0, \label{neutralino-mass}
\end{eqnarray}
where the element $N_{ij}$ parameterizes the component of the field $\psi_j^0$ in neutralino state $\tilde{\chi}_i^0$.

In supersymmetric models with R-parity conservation, the lightest neutralino $\tilde{\chi}_1^0$ acts as a promising DM candidate. Given that
 $\mu$ is usually smaller than about $300 \, {\rm GeV}$ in nNMSSM \cite{King-nNMSSM,NMSSM-125-1} and the LHC searches for electroweakinos have required
$M_2$ to be larger than about $350 \, {\rm GeV}$ in simplified scenarios \cite{3lETmiss}, one can infer that the dominant component of $\tilde{\chi}_1^0$
prefers to be any of bino, singlino and higgsinos. As has been pointed out by numerous studies, in this case
$\tilde{\chi}_1^0$ may achieve acceptable relic density in the following regions \cite{nNMSSM-LUX}
\begin{itemize}
\item Higgs boson or $Z$ boson resonance region, where the Higgs may be any of the three CP-even and two
CP-odd Higgs bosons.
\item Coannihilation region where $\tilde{\chi}_1^0$ is nearly degenerated with any of $\tilde{\chi}_1^\pm$, $\tilde{\chi}_2^0$ and $\tilde{l}$
($\tilde{l}$ represents the lightest slepton).
\item Large mixing region where $\tilde{\chi}_1^0$ has large higgsino and singlino fractions.
\end{itemize}

As for the DM physics in nNMSSM, two points should be noted. One is that since the higgsinos in nNMSSM are not heavy, i.e. $\mu \lesssim 300 \, {\rm GeV}$,
the higgsino components in $\tilde{\chi}_1^0$ are usually sizable, which can enhance the couplings of $\tilde{\chi}_1^0$ with Higgs and $Z$ bosons.
As a result, the cross sections of DM-nucleon scattering tend to be large, and thus are subject to the constraints from DM direct
detection experiments such as LUX and PandaX-II. In \cite{nNMSSM-LUX}, we have shown that such constraints are very strong in
excluding vast region in $\lambda-\kappa$ plane, which implies that the parameter region where the diphoton signal of $h_1$ is optimal will be inevitably
affected. In fact, this is one of our motivations to study the diphoton rate in light of the DM experiments.  The other point is that in most viable
case of nNMSSM, $\tilde{\chi}_1^0$ is singlino-dominated. Since the interactions of such a $\tilde{\chi}_1^0$ are rather weak,
it usually annihilated in early universe through the resonance of the singlet-dominated pseudoscalar to get acceptable
relic density. This also imposes non-trivial requirements on the parameter space of nNMSSM to affect the diphoton rate.

\section{Diphoton rate of $h_1$ in nNMSSM}

In this section, we first perform a comprehensive scan over the parameter space of the $Z_3$ NMSSM by considering various experimental constraints,
then we investigate the diphoton rate in its allowed parameter space. We present the features of the signal by deliberate figures.

\subsection{Strategy in parameter scan}

We begin our study by making some assumptions about unimportant SUSY parameters. These assumptions are consistent with
current LHC search for sparticles, and they contain following items:
\begin{itemize}
\item gluino mass and all soft breaking parameters for the first two generation squarks are set to be $2 \, {\rm TeV}$.
\item all soft parameters in third generation squark sector are treated as free parameters except
that the relations $m_{U_3} = m_{D_3}$ for right-handed soft breaking masses and $A_t = A_b$
for soft breaking trilinear coefficients are assumed for the sake of simplicity.
\item all soft breaking parameters in slepton sector take a common value $m_{\tilde{l}}$.
This quantity mainly affects the muon anomalous magnetic moment.
\end{itemize}

With the above assumptions, we use the package NMSSMTools-5.0.1 \cite{NMSSMTools} to scan the parameters
of the $Z_3$ NMSSM as follows:
\begin{eqnarray}\label{NMSSM-scan}
&& 0 <\lambda\leq 0.75,\quad  0 <\kappa \leq 0.75, \quad  2 \leq \tan{\beta} \leq 60, \nonumber \\
&&100{\rm ~GeV}\leq m_{\tilde{l}} \leq 1 {\rm ~TeV}, \quad 100 \, {\rm GeV} \leq \mu \leq 1 \, {\rm TeV}, \nonumber \\
&& 50 {\rm ~GeV}\leq M_A \leq 2 {\rm ~TeV}, \quad |A_{\kappa}| \leq 2 \, {\rm TeV}, \nonumber\\
&& 100{\rm ~GeV}\leq M_{Q_3},M_{U_3} \leq 2 {\rm ~TeV},  \nonumber \\
&& |A_{t}|\leq {\rm min}(3 \sqrt{M_{Q_3}^2 + M_{U_3}^2}, 5 \, {\rm TeV}), \nonumber\\
&& 20 \, {\rm GeV} \leq M_1 \leq 500 \, {\rm GeV}, \nonumber \\
&& 100 \, {\rm GeV} \leq M_2 \leq 1 \, {\rm TeV},
\end{eqnarray}
where all the parameters are defined at the scale of $2 \, {\rm TeV}$. To be more specific, we carry out two different sets
of Markov Chain scans to ensure our results as inclusive as possible. The first set of scans aim at getting the samples which
satisfy the experimental upper bounds on DM relic density and DM-nucleon scattering cross sections,
and the corresponding likelihood function we adopt is
\begin{eqnarray}
{\cal{L}} &=& {\cal{L}}_{m_{h_2}} \times {\cal{L}}_{Br(B \to X_s \gamma)} \times  {\cal{L}}_{Br(B_s \to \mu^+ \mu^-)} \nonumber \\
&& \times  {\cal{L}}_{\Omega h^2} \times  {\cal{L}}_{\sigma_i},
\end{eqnarray}
where ${\cal{L}}_{m_{h_2}}$, ${\cal{L}}_{Br(B \to X_s \gamma)}$ and ${\cal{L}}_{Br(B_s \to \mu^+ \mu^-)}$ are likelihood functions for
experimentally measured SM-like Higgs boson mass, $Br(B \to X_s \gamma)$ and $Br(B_s \to \mu^+ \mu^-)$ respectively, which are taken to
be Gaussian distributed, and ${\cal{L}}_{\Omega h^2}$ and  ${\cal{L}}_{\sigma_i}$ denote the likelihood functions from the upper bounds on the DM
observables with their explicit forms given in \cite{Likehood}. We select more than ten parameter points from the scan results in \cite{nNMSSM-LUX}
which are well separated in $\lambda-\kappa$ plane as the starting points of the Markov Chain scans. This set of scans, as were shown by our practices,
usually get samples with rather low $h_1$ diphoton rates. The second set of scans are designed to get the samples with a relatively large diphoton rate.
For this end, we first scan the parameter space with the likelihood function
\begin{eqnarray}
{\cal{L}} &=& {\cal{L}}_{m_{h_2}} \times {\cal{L}}_{Br(B \to X_s \gamma)} \times  {\cal{L}}_{Br(B_s \to \mu^+ \mu^-)} \times {\cal{L}_{\sigma_{\gamma \gamma}}}, \nonumber
\end{eqnarray}
where ${\cal{L}_{\sigma_{\gamma \gamma}}} = exp\left [ - \left ( \sigma^{\rm 8 TeV}_{{\rm SM}, \gamma \gamma} (h_1)/\sigma^{\rm 8 TeV}_{\gamma \gamma} (h_1) \right )^2 \right ]$
is used to look for samples with large diphoton rates. After such a preliminary scan, we obtain some representative parameter points
characterized by a large diphoton rate and meanwhile moderately large DM observables. Taking them as starting points, we then scan the parameter space of nNMSSM
again, but this time the likelihood functions for the DM observables are included. Our results indicate that such a special treatment is essential to get the
desired samples.

For the samples obtained in the scans, we further require them to explain at $2 \sigma$ level various B-physics observables, 125 {\rm GeV} Higgs boson
and muon anomalous magnetic moment, and satisfy the upper bounds set by LEP experiments, dark matter measurements as well as ATLAS analysis
on the diphoton signal of a light Higgs \cite{Aad:2014ioa}. All these quantities have been implemented in the package NMSSMTools-5.0.1.
Moreover, we impose the constraints from the direct searches for Higgs
bosons at Tevatron and LHC with the package HiggsBounds \cite{HiggsBounds}, the LHC
searches for sparticles by detailed simulation\footnote{In our previous work \cite{nNMSSM-LHC}, we introduced in detail how to
implement the direct search constraints from LHC Run-I. Here we adopt the same way as \cite{nNMSSM-LHC} to impose the constraints.},
and also the Fermi-LAT observation of dwarf galaxy \cite{Fermi}.

\begin{figure*}[t]
  \centering
  \includegraphics[height=7cm,width=15cm]{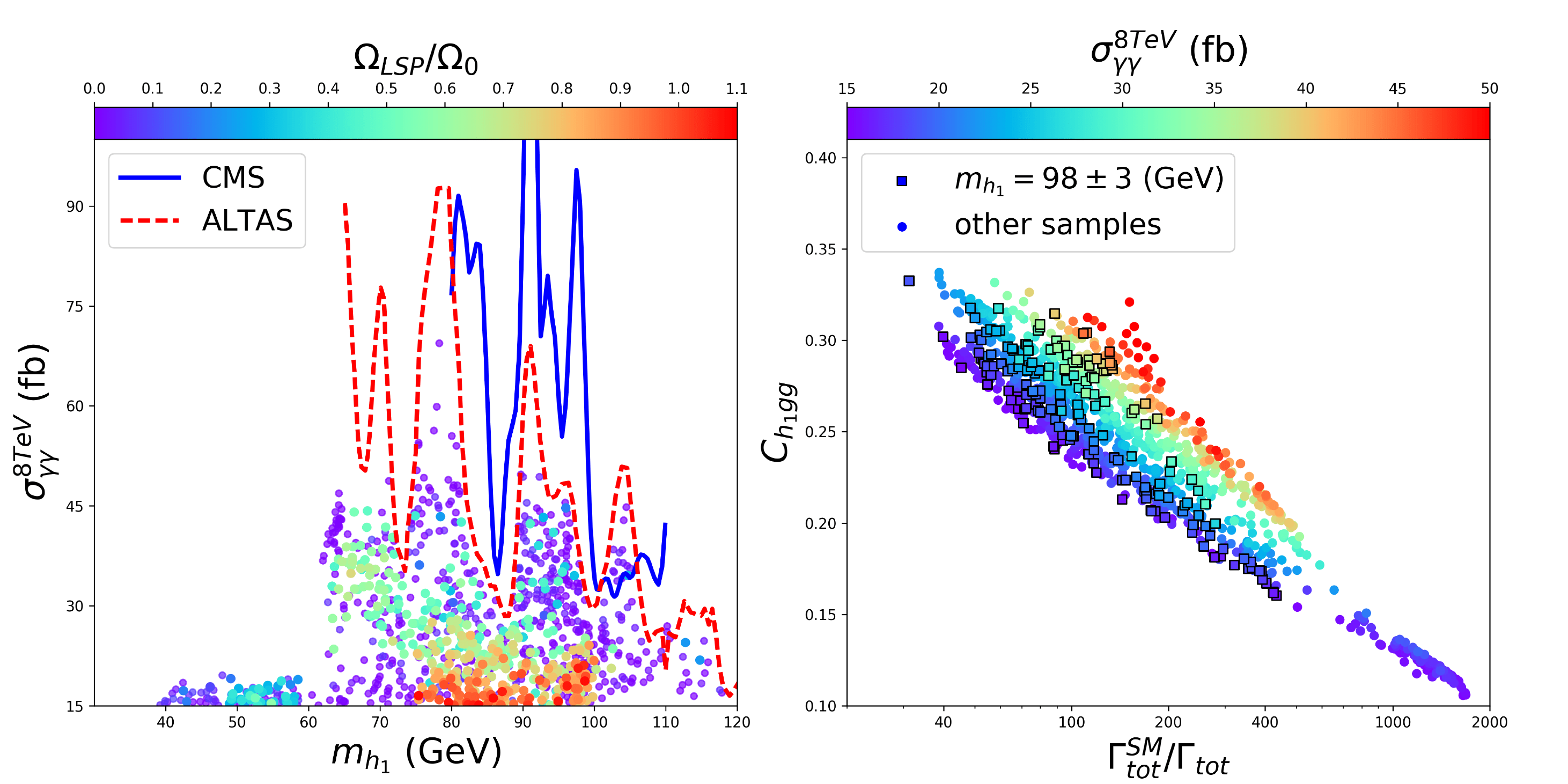} \\
  \caption{{\bf Left panel}: the diphoton rate of $h_1$ at $ 8 \, {\rm TeV}$ LHC versus $h_1$ mass for the samples surviving the constraints
  in the scan and meanwhile predicting a moderate large diphoton rate,  $\sigma^{{\rm 8 TeV}}_{\gamma \gamma} \geq 15 \, {\rm fb}$.
  Colors in this panel indicate how much $\tilde{\chi}^0_1$ constitutes the relic abundance today and the red dotted (blue solid) line
  corresponds to the current ATLAS (CMS) bounds on the rate. {\bf Right panel}: correlation of the normalized $h_1$-gluon-gluon
  coupling $C_{h_1 g g}$ to the ratio $\Gamma^{\rm SM}_{tot}/\Gamma_{tot}$ for the samples in the left panel with the colors indicating the
  magnitude of $\sigma^{8 {\rm TeV}}_{\gamma \gamma}$. $\Gamma_{tot}$ denotes the total width of $h_1$ predicted by nNMSSM, and
  $\Gamma^{\rm SM}_{tot}$ is the width of $h_1$ calculated by assuming that $h_1$ has same couplings as those of the SM Higgs boson.
  Squares in the panel represent samples with $m_{h_1}=98 \pm 3 \, {\rm GeV}$, which is the mass range favored by
  the LEP and CMS mild excesses.}\label{fig1}
\end{figure*}

The constraints we consider here differ from those of our previous works \cite{nNMSSM-LHC,nNMSSM-LUX}  in the following aspects.
\begin{itemize}
\item First, we allow for the possibility that $\tilde{\chi}_1^0$ constitutes
a fraction of DM observed in the Universe. In this case, the constraints from DM direct search experiments set an upper bound
on the weighted DM-nucleon scattering cross section $\Omega_{LSP}/\Omega_0 \times \sigma_{\tilde{\chi}_1^0-n}$ with $\Omega_{LSP} h^2$
and $\Omega_0 h^2$ denoting the relic density contributed by $\tilde{\chi}_1^0$ and the measured DM density from PLANK
\cite{relic:Planck} and WMAP 9-year data \cite{relic:WMAP} respectively.
In practice, we use the latest bounds of LUX and PandaX experiments on both spin-independent and spin-dependent
(SD) scattering rates to set limits, and since a $10 \%$ theoretical uncertainty is usually assumed in calculating
$\Omega h^2$ by the package MicrOMEGAs \cite{micrOMEGA}, we consider $\tilde{\chi}_1^0$ as the sole DM candidate
if $0.9 \leq \Omega_{LSP}/\Omega_0 \leq 1.1$.
\item Second, we consider the constraint from Fermi-LAT searches for DM-annihilation from dwarf galaxies. Since the DM annihilation
for each parameter point usually includes a variety of channels in today's Universe, which is different from those single SM final states assumed by
Fermi-LAT collaboration  to set bounds \cite{Fermi},
we actually require the $\langle \sigma v \rangle$-weighted number of photon predicted by the parameter point to be less than that calculated
with the Fermi-LAT bounds (see \cite{TopFlavoredScalarDM-Baek:2016lnv,Bringmann:2012vr,Giacchino:2015hvk} for similar usage), i.e.
$\langle \sigma v \rangle_{th} N_{\gamma,th} \lesssim \langle \sigma v \rangle_{exp} N_{\gamma,exp}$ where
\begin{eqnarray}
N_{\gamma,th} &=& \int^{E_{\gamma,max}}_{E_{\gamma,min}} dE_\gamma \frac{dN^{th}_\gamma}{dE_\gamma},  \nonumber \\
N_{\gamma,exp} &=& \int^{E_{\gamma,max}}_{E_{\gamma,min}} dE_\gamma \frac{dN^{exp}_\gamma}{dE_\gamma},
\end{eqnarray}
with $\{E_{\gamma,min}, E_{\gamma,max} \}=\{0.5,\,500\} \,{\rm GeV}$ being the photon energy range analyzed in \cite{Fermi}.
In more detail, we use the package micrOMEGAs \cite{micrOMEGA} to obtain the theoretical predictions $\langle \sigma v \rangle_{th}$ and
\begin{eqnarray}
\frac{dN^{th}_\gamma}{dE_\gamma} = \sum_f Br^{(f)} \frac{dN^{(f)}_\gamma}{dE_\gamma}.
\end{eqnarray}
We choose $\langle \sigma v \rangle_{exp} = \langle \sigma v \rangle_{b\bar{b}}$
which denotes the Fermi-LAT bound on the rate of the annihilation $\tilde{\chi}_1^0 \tilde{\chi}_1^0 \to b \bar{b}$  \cite{Fermi}. We also utilize the photon spectrum $dN^{exp}_\gamma/dE_\gamma = dN^{(b\bar{b})}_\gamma / dE_\gamma$ generated by the code
PPPC4DMID \cite{PPPC4DMID-Cirelli:2010xx}.

In order to check the validity of this simple way to implement the constraint, we alternatively use
the method proposed in \cite{Likelihood} and adopted in \cite{Zhou} to exclude parameter points.  The latter method utilizes the likelihood function
provided by Fermi-LAT collaboration \cite{Fermi-Web} and allows the variation of the $J$-factor for each dwarf galaxy. We find the two methods are
consistent as far as our samples are considered. Possible underlying reason for this is
that for the excluded samples DM annihilates mainly via the mediation of $A_1$ and consequently, the dominant final state is either
$b \bar{b}$ or $t \bar{t}$. Since the shape of the spectrum $dN^{(t\bar{t})}_\gamma / dE_\gamma$ is similar to that of
$dN^{(b\bar{b})}_\gamma / dE_\gamma$ for a given $m_{\tilde{\chi}_1^0}$, as a good approximation one may simply scale the Fermi-LAT bound
on $\langle \sigma v \rangle_{b\bar{b}}$ to get that for  $\langle \sigma v \rangle_{t\bar{t}}$ \cite{TopFlavoredScalarDM-Baek:2016lnv,Bringmann:2012vr,Giacchino:2015hvk}. Moreover, we also find that the Fermi-LAT constraint is rather weak and excludes only about 30 samples in our study. We checked that the excluded samples are featured by
$100 {\rm GeV} < m_{\tilde{\chi}_1^0} < 200 {\rm GeV}$, $0.3 < N_{13}^2 + N_{14}^2 < 0.7$, $2 m_{\tilde{\chi}_1^0} > m_{A_1}$ and
$\langle \sigma v \rangle_{\rm Today} \gtrsim 10^{-23} {\rm cm^3 s^{-1}}$.  We remind that the condition $2 m_{\tilde{\chi}_1^0} > m_{A_1}$ ensures that
the DM annihilation rate at current days is larger than that in early Universe \cite{Cao:2014efa}.
\item Third, we use the latest version of package NMSSMTools to calculate various
observables. There are many improvements of this version over previous ones, especially with the help of the
package Lilith \cite{Bernon:2015hsa} which utilizes the recently combined ATLAS and CMS analysis on $125 \, {\rm GeV}$ Higgs at LHC Run-I
\cite{Khachatryan:2016vau} to limit the model.
\item Finally, in getting the physically viable samples of nNMSSM, we do not require the fine tuning quantities $\Delta_h$ and
$\Delta_Z$ to be less than an artificial value 50 as we did
in \cite{nNMSSM-LHC,nNMSSM-LUX}, instead we only require that $h_2$ acts as the $125 \, {\rm GeV}$ Higgs boson.
\end{itemize}

\begin{table}[t]
  \caption{Ranges of some dimensional parameters and masses in unit of $\rm GeV$ obtained in the scans of this work. \label{tab1}
    }\centering

\vspace{0.2cm}

\begin{tabular}{|c|c|c|c|c|c|}
\hline
    $P_i$ & Range  & Mass & Range  & Mass & Range \\
\hline
            $M_1$           & $ 66\sim  350$  &
    $\widetilde{t}_1$       & $670 \sim 1800$  &
    $\widetilde{\chi}_1^0$  & $ 59 \sim  220$  \\
\hline
            $M_2$           & $350 \sim 860$   &
    $\widetilde{b}_1$       & $690 \sim 1850$  &
    $\widetilde{\chi}_2^0$  & $ 70 \sim 280$   \\
\hline
            $M_A$           & $485\sim1930$   &
    $\widetilde{\tau}_1$    & $ 94 \sim 670$   &
    $\widetilde{\chi}_1^\pm$  & $106 \sim 320$   \\
\hline
            $\mu$           & $104 \sim 330$   &
            $A_1$           & $ 28 \sim 420$   &
          $H^\pm_1$           & $470 \sim1950$   \\
\hline
\end{tabular}
\end{table}

In the following discussion, only the samples satisfying all of the constraints mentioned above are considered. In Table \ref{tab1},
we list the ranges of the dimensional parameters in Eq.(\ref{NMSSM-scan}) and their prediction on the mass spectrum of some particles.
Note that these sparticle spectrums are compatible with the direct searches for SUSY at LHC Run-I.

\subsection{Numerical results}

\begin{figure*}[t]
  \centering
  \includegraphics[height=5.1 cm,width=15cm]{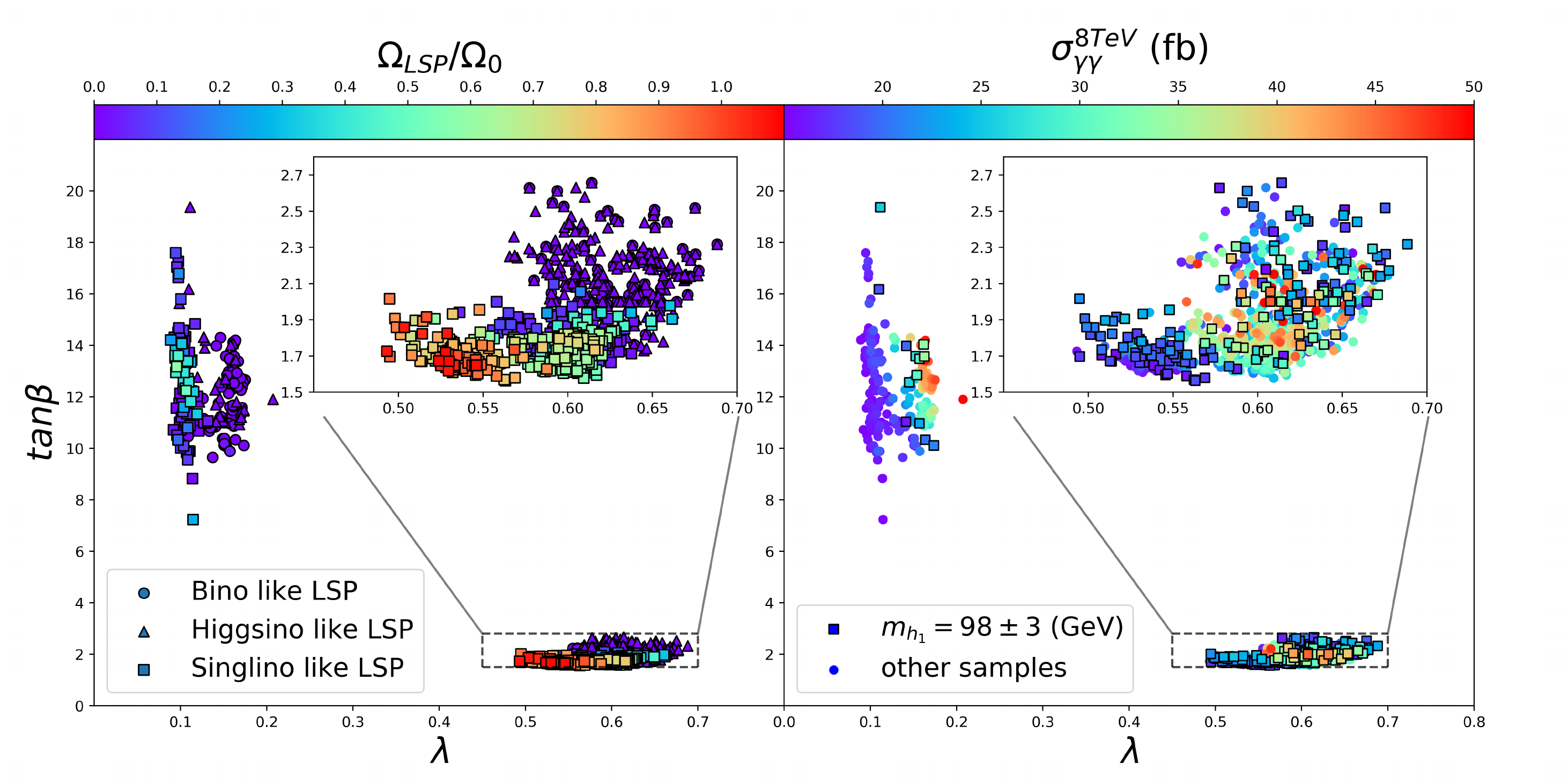} \\

  \vspace{-0.1cm}

 \includegraphics[height=4.9cm,width=15cm]{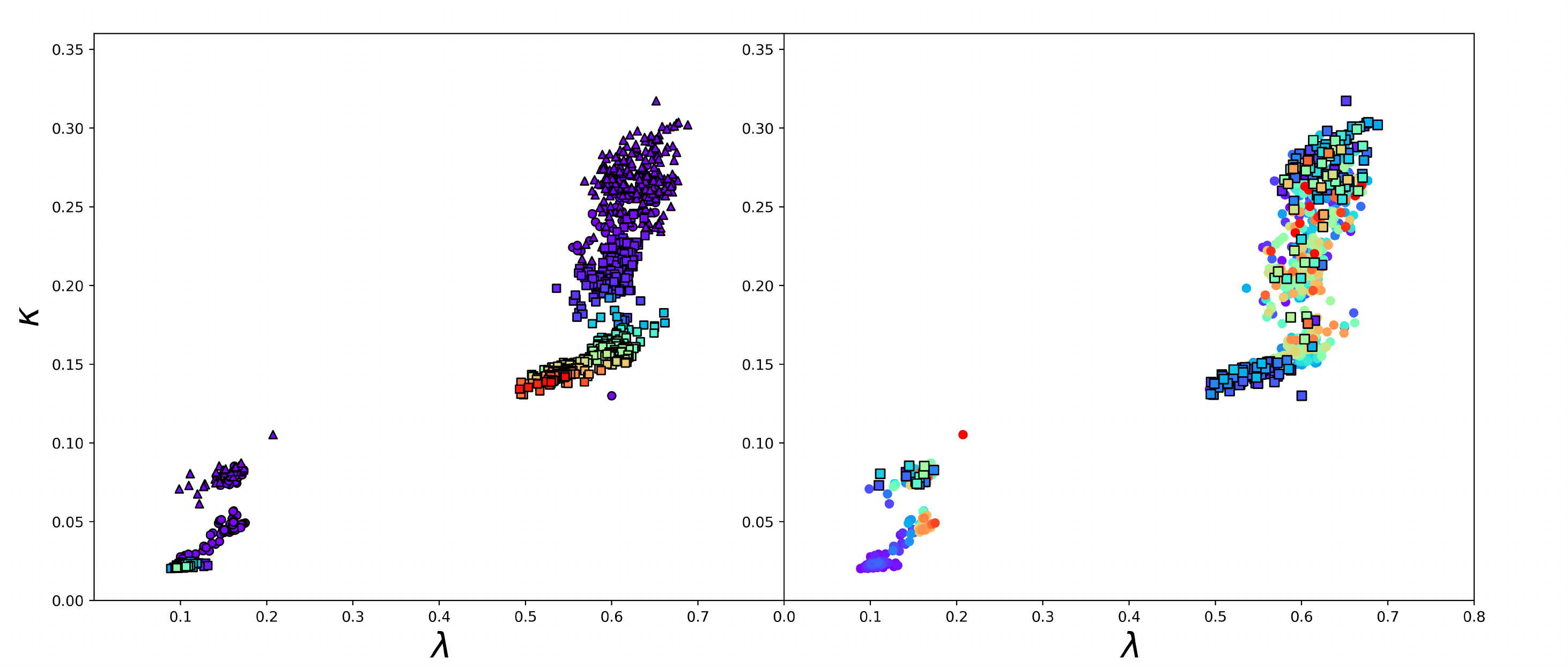} \\

 \vspace{-0.1cm}

  \includegraphics[height=4.9cm,width=15cm]{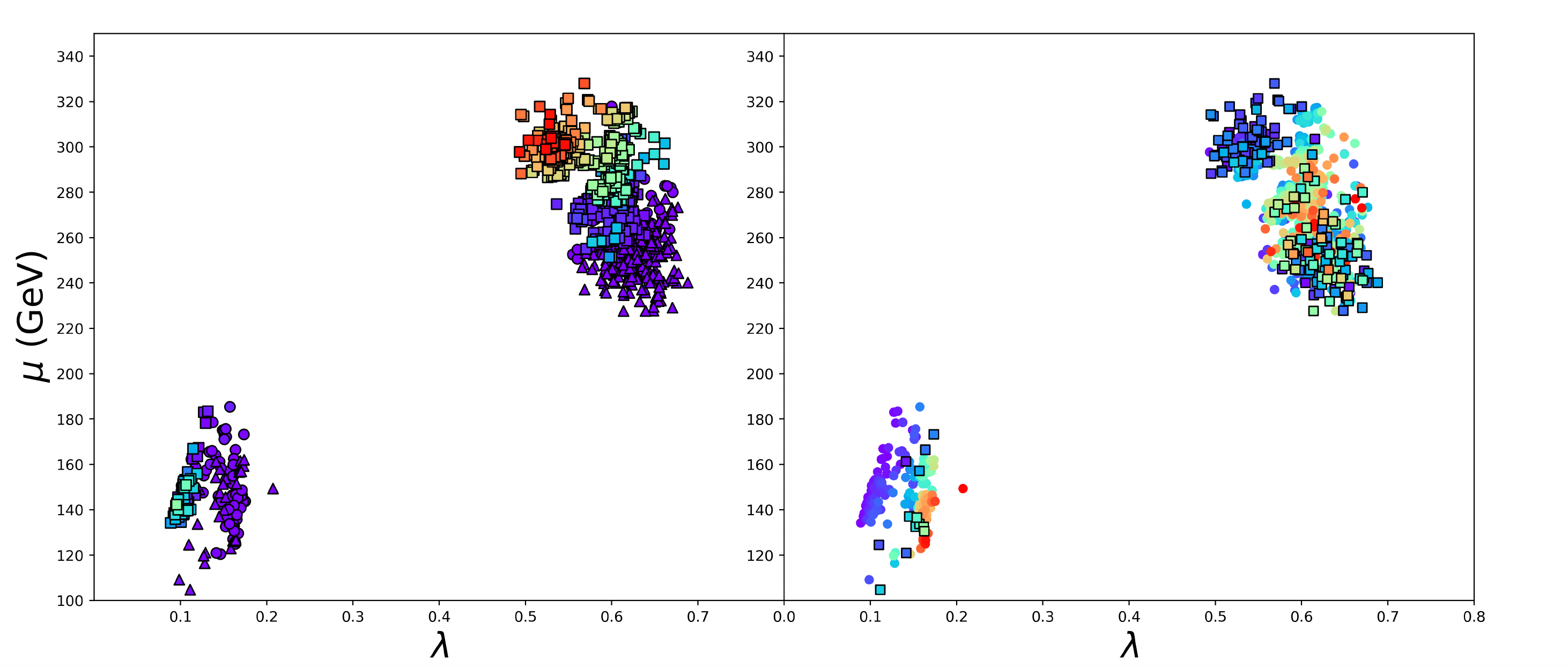} \\

  \vspace{-0.1cm}

   \includegraphics[height=4.9cm,width=15cm]{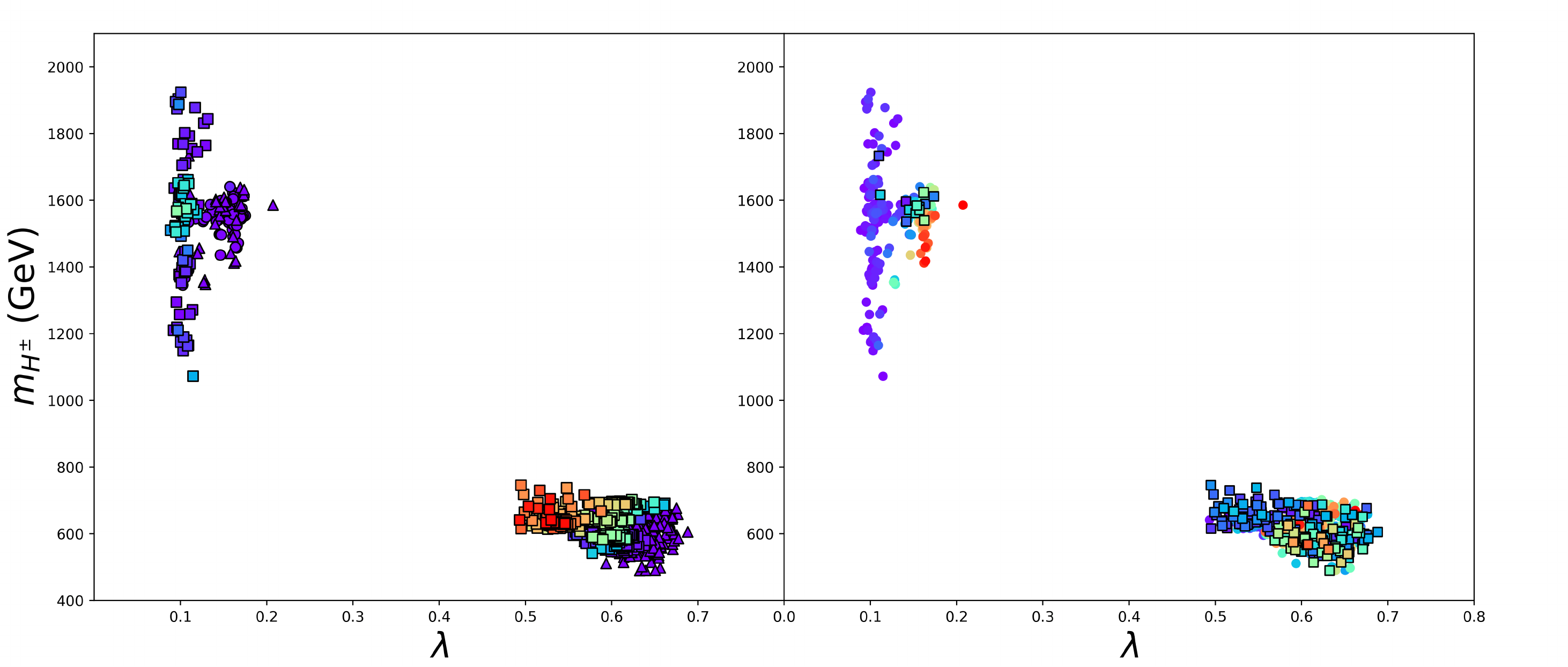}

   \vspace{-0.2cm}

\caption{Samples in Fig.\ref{fig1} projected in different parameter planes.  In left panels, colors
denote the $\tilde{\chi}^0_1$ contribution to the thermal relic $\Omega_{LSP}/\Omega_0$, and dots,
triangles and squares represent samples with bino, higgsino and singlino as the main component of $\tilde{\chi}_1^0$ respectively.
In right panels, colors represent the magnitude of $\sigma^{8 {\rm TeV}}_{\gamma \gamma}$,
 and squares correspond to the samples with $m_{h_1}=98\pm3\, {\rm GeV}$. }\label{fig2}
\end{figure*}

Since most nNMSSM samples obtained in the scans have a small diphoton rate of $h_1$ and meanwhile span a much wide parameter space,
considering all of them in discussion will make the figures presented below rather disordered, and obfuscate the main conclusions of this work.
So in this subsection we only consider those which predict $\sigma_{{\rm SUSY}}^{8 {\rm TeV}} (pp \to h_1 \to \gamma \gamma)$, hereafter
denoted as $\sigma_{\gamma \gamma}^{8 {\rm TeV}}$,
larger than 15 fb to simplify our analysis.

In the left panel of Fig.\ref{fig1} we show $\sigma^{8 {\rm TeV}}_{\gamma \gamma}$ versus $m_{h_1}$, where the colors indicate how much
$\tilde{\chi}^0_1$ constitutes the relic abundance today and the red dotted (blue solid) line corresponds to the current ATLAS
(CMS) bounds on the rate.  This figure shows that there are still plenty of nNMSSM samples which can evade current LHC
searches for a light Higgs beyond the SM, despite many of them  can not solely account for the observed relic abundance. For these samples,
the maximal prediction of the $h_1$ diphoton rate at $8 \, {\rm TeV}$ LHC is significantly smaller than the prediction without
considering the constraints from DM physics,  which was presented in \cite{Diphoton-14}.
Taking $m_{h_1}$ around 80 GeV as an example, we find that the signal rate
can reach about 70 fb if one allows $\tilde{\chi}^0_1$ to constitute only a small fraction of the thermal relic (less than $10\%$), while it drops
to about $25 \, {\rm fb}$ when the full thermal relic is required. By contrast, the $h_1$ diphoton signal rate can exceed $120 \, {\rm fb}$
if one completely ignores the DM restrictions including both thermal relic and the latest direct detection bounds \cite{Diphoton-14}. Also in some cases the DM constraints are stronger than the LHC bounds in limiting the
diphoton signal, e.g. for $m_{h_1} \simeq 80 \, {\rm GeV}$ the ATLAS analysis requires $\sigma_{\gamma \gamma}^{8 {\rm TeV}} \lesssim 90 \, {\rm fb}$ while the DM physics restricts $\sigma_{\gamma \gamma}^{8 {\rm TeV}} \leq 70 \, {\rm fb}$. 

We checked that the suppression of $h_1$ diphoton rate due to DM restrictions is general over a wide range
of $h_1$ mass, as can be seen from the sample distribution with relatively large thermal relic (warm color) at the
bottom of left panel of Fig.\ref{fig1}.
To our best knowledge, this observation has not been emphasized sufficiently before and should receive reasonable attention if one considers
the interplay between the Higgs (especially singlet extension) and DM sector in supersymmetric models. Since our original intention is to exhibit this connection in a sense as general as possible by allowing a reasonably large number of NMSSM parameters to vary in the scan, a thoroughly analytical interpretation of the $h_1$ diphoton signal suppression related to DM constraints would be very difficult and nearly impossible.
However, we can still get non-trivial hints based on two factors involved in the interplay. One is that $C_{h_1 b \bar{b}}$ coupling should be strongly suppressed in order to get an enhanced $h_1$ diphoton rate as indicated in Eq.(\ref{Diphoton-Exp-1},\ref{Diphoton-Exp-2}), which limits nNMSSM parameters to certain regions providing a proper cancellation suggested by Eq.(\ref{h1-couplings}), i.e.
\begin{eqnarray}
V_{11} \tan \beta + V_{12} \sim0,
\end{eqnarray}
where some detailed discussions about CP-even Higgs mass matrix determining rotation matrix $V$ can found, e.g. in \cite{Diphoton-8,Agashe:2012zq}.
Another one comes from the DD constraints in which the coefficients of the scalar type effective DM-quark operator used in calculating SI DM-nucleon scattering rate should be suppressed, i.e. 
\begin{eqnarray}
\frac{C_{h_1 \tilde{\chi}_1^0 \tilde{\chi}_1^0} C_{h_1 N N}}{m_{h_1}^2} + \frac{C_{h_2 \tilde{\chi}_1^0 \tilde{\chi}_1^0} C_{h_2 N N}}{m_{h_2}^2} \sim 0,
\label{DDrate}
\end{eqnarray}
where approximated formulae for $C_{h_i \tilde{\chi}_1^0 \tilde{\chi}_1^0}, C_{h_i N N}$ can be found in \cite{Badziak:2015qca,Badziak:2015exr} for DM scenarios featuring different dominant components. This requirement also puts strong constraints on the nNMSSM parameter space, especially those parameters shared in both two sectors such as $\{\lambda, \kappa, \tan\beta, \mu\}$. As a result, the $Z_3$ NMSSM compromises the two requirements and results in a moderately suppressed diphoton rate. Eq.(\ref{DDrate}) actually corresponds to a well known scenario called Blind Spots (BS) in SUSY models like MSSM and NMSSM. We refer interested readers to \cite{Badziak:2015exr,Cheung:2012qy,Huang:2014xua,Han:2016qtc} (and references therein) for more detailed discussions.

The left panel of Fig.\ref{fig1} also shows that
for $h_1$ with mass smaller than $m_{h_2}/2\approx 62$ GeV where the LHC diphoton bounds are not available,
the diphoton signal are generally below 20 fb. This suppression is due to the kinematically opening of the exotic
decay $h_{2} \to h_1 h_1$ for the SM-like Higgs boson $h_2$ which receives strong constraints from the current Higgs measurement and thus pushes $h_1$
further to the singlet component corner. Another related case of $h_3 \to h_2 h_1$ in nNMSSM can be found in \cite{Kang:2013rj}.
It should also be noted that in some other cases allowed by the DM constraints, the diphoton rates can be very close to  the current LHC diphoton bounds. With the
currently updated collision energy at 13 TeV LHC and the future high luminosity upgrade, these cases are
very likely to be discovered or excluded.

In the right panel of Fig.\ref{fig1} we show the normalized $h_1$-gluon-gluon coupling $C_{h_1 g g}$ to its SM prediction with the same Higgs mass
versus the ratio of $h_1$ total width $\Gamma^{\rm SM}_{tot}/\Gamma_{tot}$ defined in Eq.(\ref{Diphoton-Exp-1}). In this panel the colors indicate the magnitude of $\sigma^{8 {\rm TeV}}_{\gamma \gamma}$ and the squares correspond to samples with $m_{h_1}=98 \pm 3 \, {\rm GeV}$ which is the
mass range favored by the LEP and CMS diphoton mild excesses. One can learn that although
the singlet-dominant nature of $h_1$ causes an overall suppression of its couplings to the SM fermions and thus to
the gluons via the fermion loop,  $C_{h_1 g g}$ can still reach about 0.35 which
is crucial to obtain a sizable $h_1$ production cross section. On the other hand, a significant suppression of $h_1$
total width compared to its SM prediction\footnote{In the following when we use the phrase 'its SM prediction', we
mean the case where $h_1$ is identical to the Higgs boson in the SM except that its mass is adopted same as the
prediction of the NMSSM.} is also needed to increase the diphoton rate
as indicated by Eq.(\ref{Diphoton-Exp-1}). This is the natural consequence of the dominant singlet component in
$h_1$ which reduces the leading decay modes into $b\bar{b},\tau^+\tau^-$.
We checked that for the samples with $\sigma^{8 {\rm TeV}}_{\gamma \gamma}$ around $30 \, {\rm fb}$, $Br(h_1 \to b\bar{b})$ is usually below
$30\%$ compared to about $90\%$ for its SM prediction, and $Br(h_1 \to \gamma \gamma)$ can reach $3\%$.

As mentioned in Section I, the existence of a light $h_1$ is tightly limited not only from the LEP measurements but also from the DM
observations. To pass the current stringent bounds from LUX and
PandaX-II experiments, there must exist strong cancelations among the contributions of the three CP-even Higgs bosons,
which would limit the nNMSSM parameter space into certain regions. In order to illustrate this expectation, in Fig.\ref{fig2}
we project the samples in Fig.\ref{fig1} on $\tan \beta-\lambda$ planes (first row), $\kappa-\lambda$ planes (second row),
$\mu-\lambda$ planes (third row) and $m_{H^\pm}-\lambda$
(last row) with the colors in left panels denoting  the $\tilde{\chi}^0_1$ contribution to the thermal relic $\Omega_{LSP}/\Omega_0$
and those in right panels representing the magnitude of $\sigma^{8 {\rm TeV}}_{\gamma \gamma}$. Moreover, we also use dots, triangles and
squares in the left panels to denote samples with bino, higgsino and singlino as the dominant component of $\tilde{\chi}_1^0$
respectively, and squares in the right panels
to denote samples with $m_{h_1}=98\pm3\, {\rm GeV}$. Obviously, given the horizontal axis
assigned to singlet-doublet-doublet Higgs coupling coefficient $\lambda$ for all panels, the samples
only moves vertically between panels with different paired nNMSSM parameters.

Fig.\ref{fig2} indicates that the samples in Fig.\ref{fig1} are distributed in two isolated parameter regions, which are given by
\begin{itemize}
\item Region I: \quad $0.1 \lesssim \lambda \lesssim 0.2$, \quad $6 \lesssim \tan \beta \lesssim 20 $, \quad $0.02 \lesssim \kappa \lesssim 0.1 $,
      \quad $100 \, {\rm GeV} \lesssim \mu \lesssim 190 \, {\rm GeV} $, \quad $1 \, {\rm TeV} \lesssim m_{H^\pm} \lesssim 2 \, {\rm TeV} $;
\item Region II: \quad $0.45 \lesssim \lambda \lesssim 0.70$, \quad $1.5 \lesssim \tan \beta \lesssim 3 $, \quad $0.1 \lesssim \kappa \lesssim 0.3 $,
    \quad $220 \, {\rm GeV} \lesssim \mu \lesssim 330 \, {\rm GeV} $, \quad $450 \, {\rm GeV} \lesssim m_{H^\pm} \lesssim 700 \, {\rm GeV} $.
\end{itemize}
Since the colors in the left and right panels correspond to $\Omega_{LSP}/\Omega_0$ and $\sigma^{8 {\rm TeV}}_{\gamma \gamma}$ respectively, one can
quickly identify that only part of samples in Region II can have $\tilde{\chi}^0_1$ capable of accounting for all of the DM relic density today\footnote{We emphasize that only samples with $\sigma^{8 {\rm TeV}}_{\gamma \gamma} \geq 15 \, {\rm fb}$ are shown in Fig.\ref{fig2}.
If we do not consider such a requirement, the parameter $\lambda$ for experimentally allowed samples will span a wide range from
$0.03$ to $0.7$, and $\tilde{\chi}_1^0$ can account for the measured relic density at any value of $\lambda$ \cite{nNMSSM-LUX}.
We obtained this observation by intensive and time-consuming scans. During the process we also noticed that it was rather difficult to obtain nNMSSM samples satisfying all the constraints, especially when one requires $\tilde{\chi}_1^0$ to fully
account for the relic density. This reflects the fact that parameters closely related to DM properties,
such as $\lambda$, $\kappa$, $\tan \beta$ and $\mu$, must collaborate properly to survive the constraints.}.
For these samples, $\tilde{\chi}_1^0$ is singlino-dominated which can seen
clearly from the enlarged region in the first row of Fig.\ref{fig2}. This can also be inferred from
the relation $2 \kappa/\lambda < 1$ as shown in the second row of diagrams. We checked that $\tilde{\chi}^0_1$ for this
case annihilated in early universe mainly through the $s$-channel exchange of a moderately light singlet-like
$A_1$ to get an acceptable relic density.

As for Region II, one should note that the charged Higgs boson is moderately light and consequently, $Br^{(th)}( B \to X_s \gamma)$ may
deviate significantly from its SM prediction. We checked that the ratio varies from
$3.75 \times 10^{-4}$ to $ 4.2 \times 10^{-4}$ (In NMSSMTools, the theoretical uncertainties is included in the calculation. So the central value of $Br^{(th)}(B \to X_s \gamma)$ is allowed to vary in
a broader range than its experimentally favored range). We also checked that in this region $H^\pm$ is approximately degenerated in 
mass with $H_3^0$ and the doublet-dominated CP-odd Higgs boson. In our analysis, we have included the constraint on the neutral sector 
from the LHC direct searches for extra Higgs bosons in terms of $\tau \bar{\tau}$ final state \cite{CMS-Extra-Higgs} 
through both the package NMSSMTools and the package HiggsBounds.

\begin{figure*}[t]
  \centering
  \includegraphics[height=8cm,width=15cm]{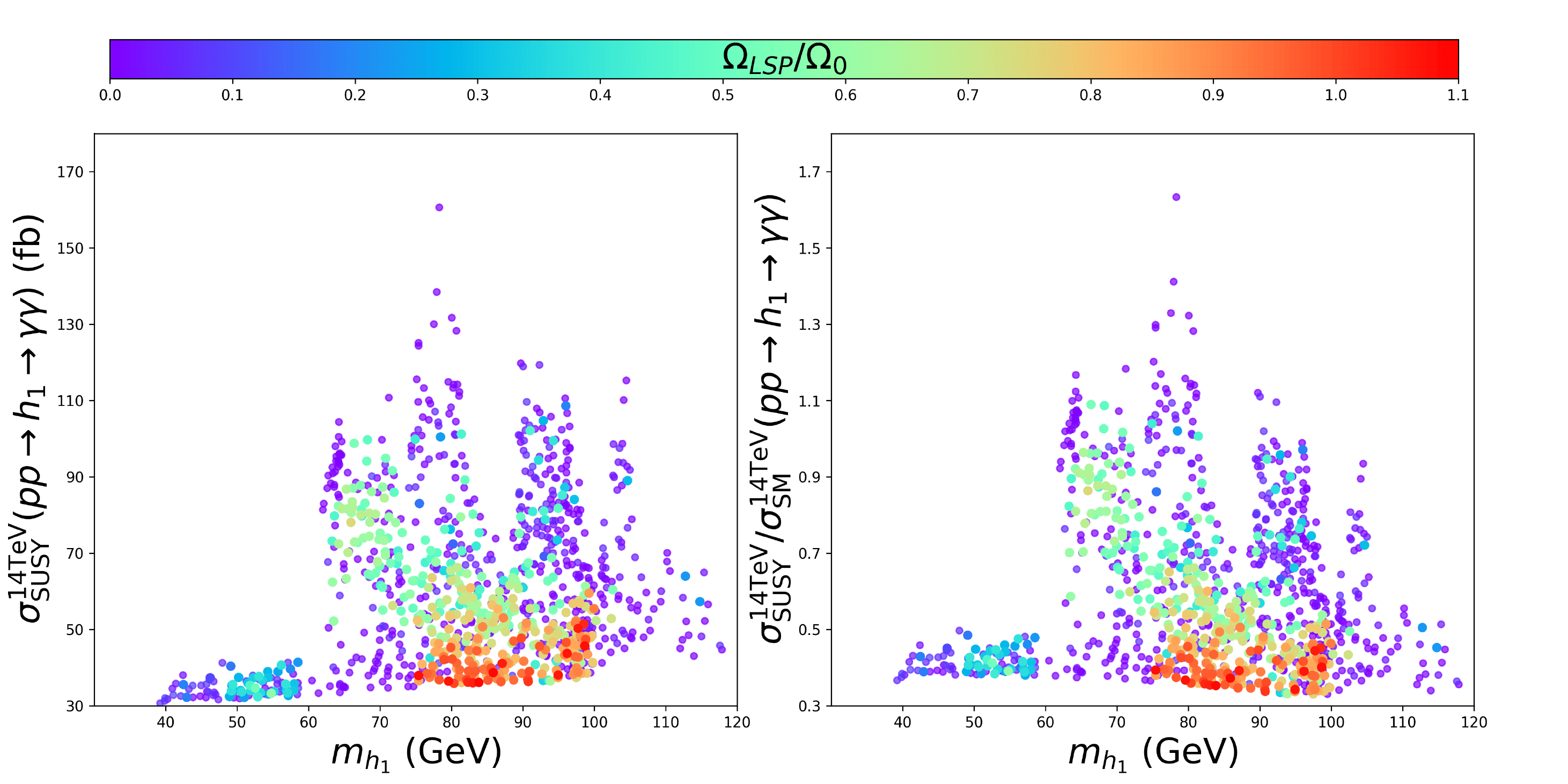} \\
  \caption{Same as the left panel in Fig.\ref{fig1} except that the vertical axes denote the diphoton rate at 14 TeV LHC.
  In the right panel, the ratio $\sigma_{\rm SUSY}^{\rm 14 TeV}/\sigma_{\rm SM}^{\rm 14 TeV}$ represents normalized diphoton rate
  where the cross section $\sigma^{\rm 14 TeV}_{\rm SM} ( p p \to h_1 \to \gamma \gamma )$ is calculated by assuming that $h_1$
  has the same couplings as those of the SM Higgs boson. Note that this normalized signal rates are independent of LHC collision energy in our case where the gluon fusion dominates the $h_1$ production.}  \label{fig3}
\end{figure*}

\begin{figure*}[t]
  \centering
  \includegraphics[height=8cm,width=15cm]{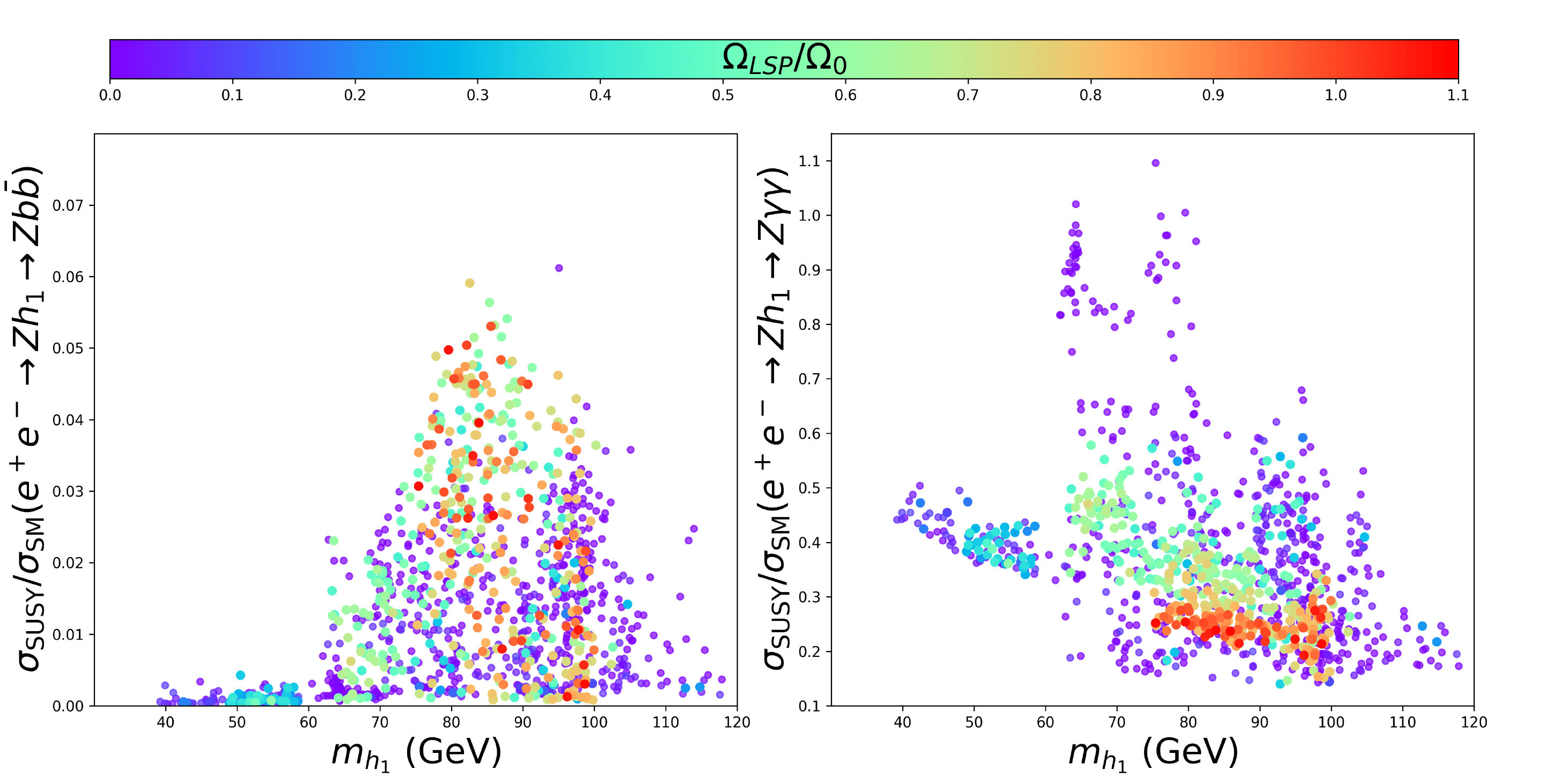} \\
  \caption{Similar to the right panel of Fig.\ref{fig3}, but displaying the normalized rate for the process $e^+ e^- \to Z h_1 \to Z b \bar{b}$ (left panel) and  $e^+ e^- \to Z h_1 \to Z \gamma \gamma$ (right panel). Note that these normalized rates are independent of $e^+ e^-$ collision energy. }\label{fig4}
\end{figure*}

\begin{figure*}[t]
  \centering
 \hspace*{-0.7cm} \includegraphics[height=11cm, width=15cm]{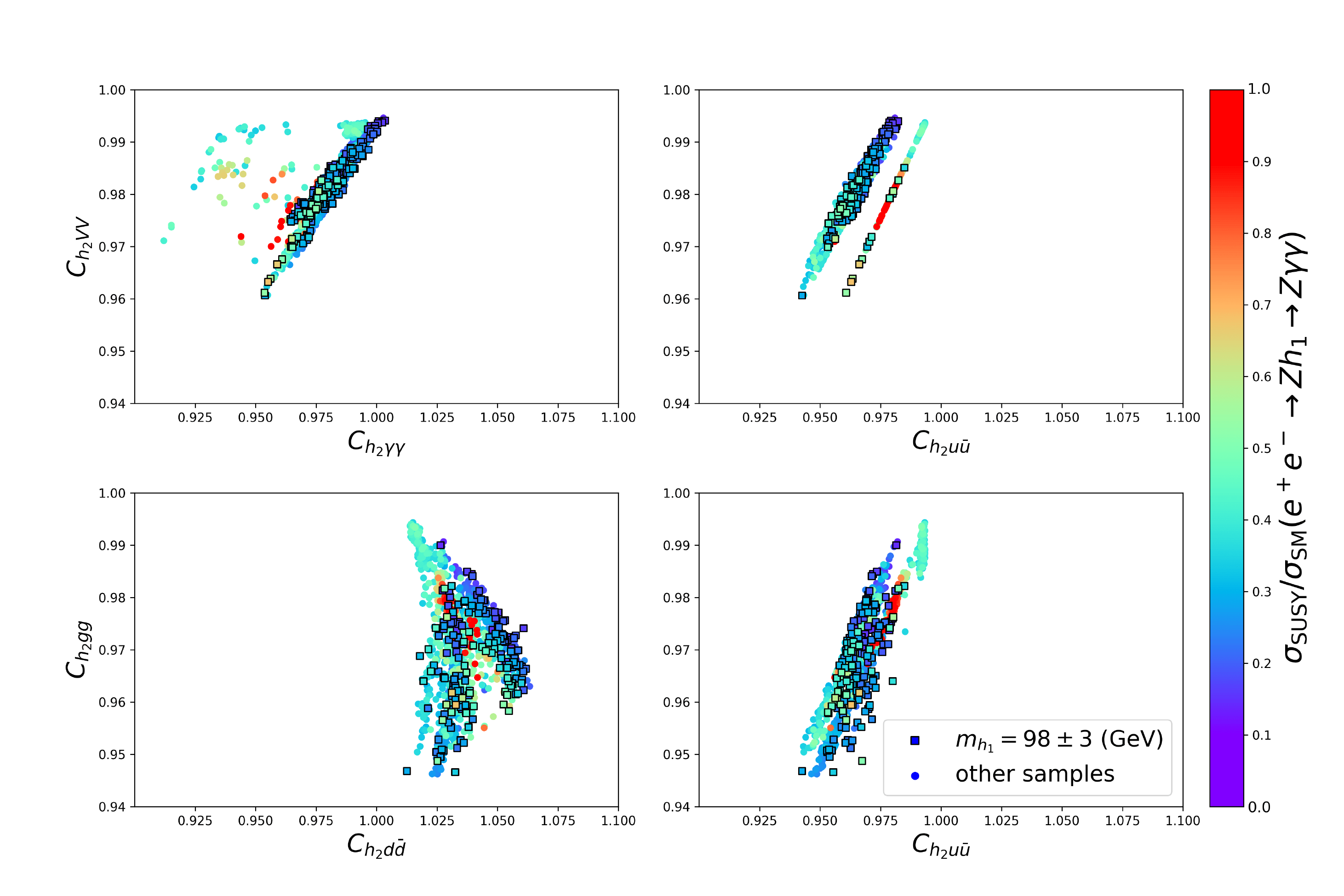} \\
  \caption{Normalized couplings of the SM-like Higgs boson $h_2$ for the samples in Fig.\ref{fig1} with colors denoting the normalized
  diphoton rate at futuer $e^+ e^-$ collider. This figure reflects the correlation of the $h_2$ couplings with the $h_1$ diphoton rate
  at $e^+ e^-$ collider. }\label{fig5}
\end{figure*}

To gain a sense of future detection potential of $h_1$ via the diphoton signal, in the left panel of Fig.\ref{fig3} we show the diphoton rate
$\sigma^{14 {\rm TeV}}_{\gamma \gamma}$ versus $m_{h_1}$, which is similar to Fig.\ref{fig1} but with $\sqrt{s}=14$ TeV at the LHC.
One can learn that a general cross section enhancement of $2 \sim 3$ times can be achieved with the increased collision energy, e.g.
for $m_{h_1}$ around 80 GeV the diphoton signal rate can reach about 160 fb instead of 70 fb at 8 TeV LHC. On the right
panel of Fig.\ref{fig3} we further compare the $h_1$ diphoton rate to its SM prediction. We can see that despite the general suppression
of the $h_1$ couplings to SM particles, an increased diphoton signal as large as 1.6 times can still be achievable in the light
Higgs mass region due to the suppression of the $h_1$ total width. Note that LHC as a hadron collider suffers from large hadronic background,
and consequently the diphoton signal is usually the most ideal channel to search for $h_1$ in spite of the fact that $b\bar{b}$ is generally the
dominant decay mode of $h_1$. If the diphoton signal is discovered in future with a moderately large rate, Fig.\ref{fig3} can provide us useful information about whether $\tilde{\chi}_1^0$ in the nNMSSM is capable of explaining all the DM density.

Since future $e^+ e^-$ collider like Higgs factory TLEP \cite{TLEP-1,TLEP-2} and CEPC \cite{CEPC} is very powerful in discovering possibly
new light Higgs, we study the process $e^+e^- \to Zh_1$ followed by $h_1 \to b\bar{b},\gamma\gamma$. In Fig.\ref{fig4}, we show
the production rates of the two signals for the samples in Fig.\ref{fig1}. We present our results in term of the ratio of the
rate to its SM prediction which we would call normalized signal rate hereafter. Note that these normalized signal rates
are independent of the collision energy. The left panel indicates
that the $b\bar{b}$ signal rate of $h_1$ is usually strongly suppressed in comparison with its SM prediction, reaching
at most $7\%$ for the samples we considered. By contrast,  the $\gamma\gamma$ signals have a signal ratio
from mild suppression to an enhancement of 1.1 as indicated by the right panel. In order to estimate the sensitivity of
the collider to the signals, we recall that the expected precision of determining the $b \bar{b}$ signal of the SM Higgs boson is
around $0.1\%$ for TLEP \cite{Higgs-Factory} (due to the large production rate of the signal as well as the
clean background of the collider), and that for the diphoton signal is at $3\%$ level. So if we
assume the sensitivities to detect $h_1$ signals to be at the same order as those of the $125 \, {\rm GeV}$ Higgs boson, we can expect
that most samples considered in this section have an opportunity of being explored by both the $b\bar{b}$ signal and the diphoton signal at TLEP.

Apart from the direct searches for $h_1$, one can also constrain the nNMSSM parameter space via its correlation with
the properties of the SM-like Higgs boson which will be measured to a high precision at future $e^+ e^-$ collider
(about $1.5\%$ for $h_2 \gamma \gamma$ coupling and $0.5\%$ for the other couplings at TLEP
\cite{Higgs-Factory}). In Fig.\ref{fig5} we show
various couplings of $h_2$ normalized to its SM value with the colors indicating the normalized rate
for the process $e^+e^- \to Zh_1 \to Z \gamma\gamma$ to its SM prediction.
Again, we use the  squares to denote the samples with $m_{h_1}=98\pm3 \,{\rm GeV}$.
From the figure it is obvious that if future Higgs precision measurement limits the normalized couplings
within certain narrow regions, lots of currently available nNMSSM samples will be excluded and the properties of $h_1$
will be further limited. This fact implies that the precision measurement of the $h_2$ couplings plays a complementary role
to the direct searches for the light Higgs $h_1$ at the $e^+ e^-$ collider. Moreover, since the two methods are independent,
they can be used to crosscheck wether the NMSSM is the right underlying theory for the light Higgs boson once the existence of $h_1$ is
confirmed in experiment.

Before we end this section, we have the following comments about our study:
\begin{itemize}
\item From previous description, it is obvious that we actually repeated the work \cite{Diphoton-14}, where the constraints from DM physics on nNMSSM were neglected. We found
that after including the constraints, more than $90\%$ samples in our repetition were excluded and the allowed parameter
region and the diphoton rate were affected significantly. We thank the authors of \cite{Diphoton-14} for providing benchmark points
in their work for comparison.
\item In order to crosscheck our results presented in this section, we also performed same parameter scan by the package
SARAH \cite{Staub:2013tta} which employs the code SPheno \cite{Porod:2003um} as a spectrum generator. We found that
we can reproduce the results obtained by NMSSMTools except that a longer time is needed in calculation.
\item The conclusion that the diphoton rate is strongly limited after considering the DM constraints may not be applied directly to
other extensions of the $Z_3$ NMSSM. For example, in the general NMSSM model more free parameters enter the mass matrix for CP-even Higgs
bosons and also that for neutralinos \cite{NMSSM-Report}. Consequently, the parameter space which predicts a suppressed $h_1 b \bar{b}$ coupling may still be compatible with
DM observations and thus allow for an enhanced diphoton rate.  Detailed analysis of this situation is beyond the scope of our work. 
Another example is the case in which $Z_3$ NMSSM is embedded in gauge
mediated SUSY breaking framework  (GMSB). In this scenario, light gravitino usually acts as DM candidate (see \cite{Martin:1997ns,Kolda:1997wt}
for reviews and \cite{Hamaguchi:2014sea} for recent attempts)
and it can achieve correct relic density from a proper reheating history after inflation \cite{Krauss:1983ik,Nanopoulos:1983up}
and/or from NLSP decays \cite{Feng:2003xh,Feng:2003uy,Ellis:2003dn,Feng:2004zu}. Meanwhile, due to its lightness and very weak couplings,
the gravitino DM is easy to evade current and future direct detection bounds. Since the DM physics is quite different from that of the $Z_3$
NMSSM discussed in this work, its interplay with the diphoton rate should be very weak.

\end{itemize}

\section{Explanation of 98 GeV excesses in nNMSSM}

\begin{figure*}[t]
  \centering
  \includegraphics[height=8cm, width=10cm]{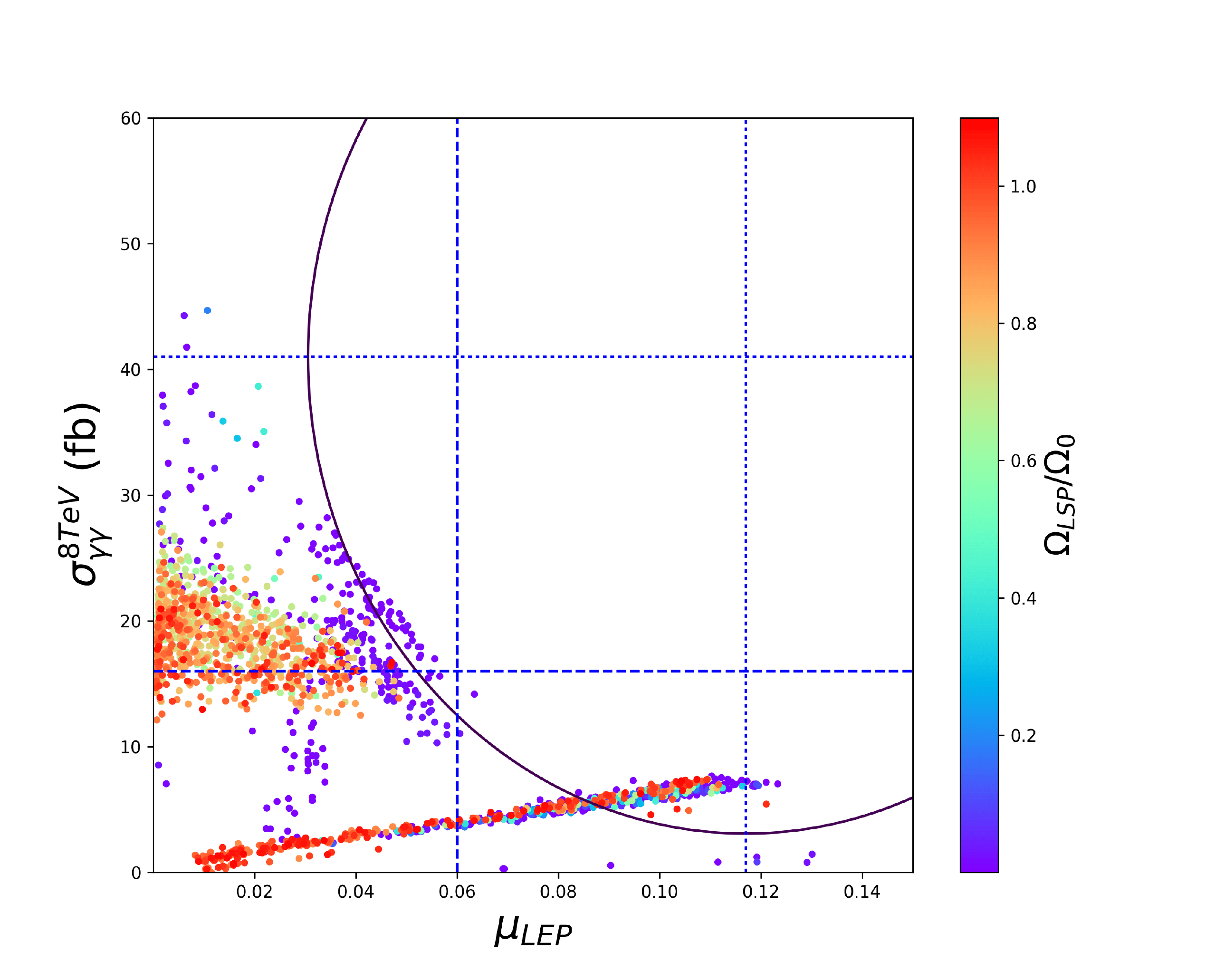} \\
  \caption{nNMSSM explanation of the excesses observed by the LEP and CMS experiments where the colors denote
  the fraction of $\tilde{\chi}_1^0$ constituing the total DM.  The horizontal and vertical blue dotted lines represent
 the central values of the two excesses respectively, and the dashed lines are their $1\sigma$ lower bounds. The
 boundary of the $1\sigma$ region for the excesses is also plotted as blue solid line. }\label{fig6}
\end{figure*}

In this section, we investigate whether nNMSSM can explain simultaneously the $98 \, {\rm GeV}$ excesses observed by both LEP and CMS experiments. For this end,
we first extract the favored signal rates from the 95\% C.L. expected and observed exclusion limits in \cite{Barate:2003sz,CMS-HIG-14-037}
with the method introduced in
\cite{Azatov:2012bz}, which are
\begin{eqnarray}
\hat{\mu}_{{\rm LEP}} =0.117\pm0.057, \quad \quad \hat{\sigma}_{\gamma \gamma}^{\rm 8 TeV} =41 \pm 25 \ {\rm fb}.  \label{central-value}
\end{eqnarray}
Then we build the following $\chi^2$
\begin{eqnarray}
\chi^2 =  \frac{(\mu_{\rm LEP} - 0.117)^2}{0.057^2}+\frac{(\sigma^{\rm 8 {\rm TeV}}_{\gamma \gamma}  -41)^2}{25^2}
\end{eqnarray}
to fit the excesses with the diphoton cross section $\sigma_{\gamma \gamma}^{\rm 8 {\rm TeV}} $ in unit of ${\rm fb}$.
In Eq.(\ref{central-value}), the first number on the right side of each formula denotes the central value of the corresponding $h_1$ signal, and
the second number is the experimental uncertainty. The quantity $\mu_{{\rm LEP}}$ is defined by
\begin{eqnarray}
\mu_{{\rm LEP}}  = \frac{\sigma_{\rm NP} (e^+e^-\to Zh_1)}{\sigma_{\rm SM} (e^+e^-\to Zh_1)} {\rm BR}(h_1\to b\bar{b}),
\end{eqnarray}
where $\sigma_{\rm NP} (e^+e^-\to Zh_1)$ denotes new physics prediction on the cross section of the process $e^+ e^- \to Z h_1$ at LEP-II.

In order to study the excesses in the framework of nNMSSM, we select some samples obtained in the scan with $m_{h_1} = 98 \pm 3 \, {\rm GeV}$ (here $3 \, {\rm GeV}$
represents the theoretical uncertainty of $m_{h_1}$), and project them on $\sigma_{\gamma \gamma}^{8 {\rm TeV}}-\mu_{\rm LEP}$ plane.
The results are given in Fig.\ref{fig6}, where
the colors indicate how much $\tilde{\chi}_1^0$ constitutes the relic abundance today. The horizontal and vertical blue dotted lines represent
the central values of the two excesses respectively, and the dashed lines are their $1\sigma$ lower bounds. We also plot
the boundary of the $1\sigma$ region favored by the excesses (blue solid line), which corresponds to $\chi^2 = 2.3 $ for two degree of freedom.
From the figure, one can learn that in nNMSSM it is very difficult to produce the central values of the two excesses simultaneously, even though the central
value of each excess can be reproduced separately and there
exist lots of samples which can explain the excesses at $1 \sigma$ level. We checked that two reasons can account for this conclusion. On the one hand, as we
introduced in Section II,  a large diphoton rate at the LHC needs a suppression of $Br(h_1 \to b \bar{b})$ and thus a suppressed $\mu_{\rm LEP}$.
On the other hand, since the property of $h_1$ is correlated with that of the SM-like Higgs boson $h_2$, the constraints on the properties of $h_2$ from relevant LHC data forbid
the associated existence of a large $\sigma_{\gamma \gamma}^{\rm 8 TeV}$ with a moderately large $\mu_{\rm LEP}$.

\begin{table*}[t]
\centering
\caption{Detailed information of four benchmark points for the $98 \, {\rm GeV}$ excesses. These samples are take from Fig.\ref{fig6} with the $\chi^2$
as low as possible.} \label{table-1}

\vspace{0.3cm}

\begin{tabular}{|l|c|c|c|c|c|c|c|c|c|c|c|c|c|}
\hline
  & $\mu_{\rm LEP}$ & $\sigma_{\gamma\gamma}^{\rm 8 TeV}$ & $\Omega_{LSP} h^2$ & $\lambda$ & $\kappa$ & $tan\beta$ & $M_A$  & $m_{h_1}$ & $C_{h_1tt}$ & $C_{h_1bb}$ & $C_{h_1\gamma\gamma}$ & $C_{h_1gg}$ & $C_{h_1VV}$ \\ \hline
P1 & 0.042                  & 19.9                    & 0.112        & 0.566     & 0.142    & 1.8        & 714.6  & 96.4      & 0.092       & 0.017     & 0.106                 & 0.099       & 0.069       \\ \hline
P2 & 0.063                  & 14.2                    & 0.001        & 0.124     & 0.059    & 12.7       & 1641.1 & 98.1      & 0.092       & 0.029     & 0.110                 & 0.098       & 0.092       \\ \hline
P3 & 0.110                  & 7.4                     & 0.110        & 0.030     & 0.014    & 22.9       & 1638.4 & 100.1     & 0.136       & 0.120     & 0.140                 & 0.138       & 0.136       \\ \hline
P4 & 0.115                  & 7.6                     & 0.001        & 0.028     & 0.009    & 17.0       & 1615.9 & 99.7      & 0.141       & 0.128     & 0.147                 & 0.144       & 0.141       \\ \hline
\end{tabular}
\end{table*}

Fig.\ref{fig6} also indicates that the samples with a low $\chi^2$ can be classified into two categories by the value of $\sigma_{\gamma \gamma}^{8 {\rm TeV}}$ and $\mu_{\rm LEP}$, which are
\begin{itemize}
\item Solution I: samples with $\sigma_{\gamma \gamma}^{8 {\rm TeV}} \gtrsim 15 \, {\rm fb}$ and $\mu_{\rm LEP} \lesssim 0.06$ (see discussion in Section 3);
\item Solution II: samples with $\sigma_{\gamma \gamma}^{8 {\rm TeV}} \lesssim 10 \, {\rm fb}$ and $\mu_{\rm LEP} \gtrsim 0.06$.
\end{itemize}
In Table \ref{table-1}, we list detailed information of four benchmark points for the excesses. Points P1 and P2 belong to Solution I and they predict $\chi^2 = 2.44$,
$\Omega_{LSP}/\Omega_0 \simeq 1$ and  $\chi^2 = 2.05 $, $\Omega_{LSP}/\Omega_0 \ll 1$ respectively. For these two points, $V_{11} \tan \beta + V_{12}$
in Eq.(\ref{h1-couplings}) is more suppressed so that the normalized coupling $C_{h_1 b \bar{b}}$ is significantly smaller than the other couplings. However,
a slight difference between the two points comes from the mass scale of the new Higgs doublet field $m_A$. Point P1 corresponds to a relatively small $m_A$ which usually implies a moderately large $V_{11}$. In this case a small $\tan \beta$ is needed for the cancelation between $V_{11} \tan \beta$ and $V_{12}$. On the contrary, point P2 predicts a large $m_A$ and thus a small $V_{11}$, in which case a large $\tan \beta$ is necessary for the cancelation.
Points P3 and P4 belong to Solution II and they have $\chi^2 = 1.78$, $\Omega_{LSP}/\Omega_0\simeq 1$ and  $\chi^2 = 1.81 $, $\Omega_{LSP} /\Omega_0  \ll 1$
respectively. These two points are characterized by $V_{11} \simeq 0$ and as a result all normalized couplings of $h_1$ are roughly equal.
In this case, both the $b\bar{b}$ and $\gamma\gamma$ signal rate can be obtained from their SM predictions by multiplying the square of
the common suppression factor for the couplings.

Finally, we emphasize that so far point P3 can explain the excesses in the best way, and at same time predicts the right relic density of DM.
For this point, the $b\bar{b}$ signal rate is around the central value of the $Z b \bar{b}$ excess while the $\gamma\gamma$ rate is somewhat small
and just around $7 \, {\rm fb}$.  On the other hand, this point is at the edge of being excluded by current LHC data of the SM-like Higgs boson,
which implies a potential tension of the LEP excess with the $125 \, {\rm GeV}$ Higgs data.

\section{Conclusion}

As an attractive scenario, natural NMSSM (nNMSSM) can predict one CP-even Higgs boson
satisfying $m_{h_1} \lesssim 120 \, {\rm GeV}$ and Higgsinos lighter than about 300 GeV. Consequently
the cross section for DM-nucleon scattering in this scenario is usually quite large, which implies that it
will be tightly limited by the recent results of LUX and PandaX-II experiments. In this work, we first
scan the parameter space of nNMSSM by considering various experimental constraints systematically. One main improvement of our study over
previous ones is that we allowed the possibility of multiple DM candidates in the Universe by not
requiring $\tilde{\chi}_1^0$ to be responsible for all of the measured DM relic density.
We find that even with such a relaxed condition, the constraint from DM physics is still strong.

Next we considered the effect of DM physics on the diphoton rate of the light Higgs. We find that the optimal value of the
signal rate at 8 TeV LHC is greatly reduced in comparison with earlier predictions.  Taking $m_{h_1}$ around 80 GeV as
an example, the signal rate can reach about 70 fb if one allows $\tilde{\chi}^0_1$ to constitute only a small
fraction of the thermal relic (less than $10\%$), and it drops to about $25 \, {\rm fb}$ when the full thermal relic is
required. By contrast, the $h_1$ diphoton signal rate can exceed $120 \, {\rm fb}$
if one completely ignores the DM restrictions. We also
briefly studied the detection potential of the light Higgs via the diphoton signal at future LHC and Higgs factory,
and observed that they have a good chance of exploring some parameter space of nNMSSM.

Finally, we investigated to what extent nNMSSM can explain the $98 \, {\rm GeV}$ excesses observed by both LEP and
CMS experiments. We conclude that there exist lots of samples which can explain the excesses at $1 \sigma$ level, even
though nNMSSM can not produce the central values of the two excesses simultaneously. The most favored samples of nNMSSM
predict the central value of the $Z b \bar{b}$ excess at LEP and a light Higgs diphoton rate at about $7 \, {\rm fb}$.

\section*{Acknowledgement}

We thank Prof. Ulrich Ellwanger and Dr. Mat$\acute{i}$as V$\acute{a}$zquez for helpful discussion about their works on
diphoton rate,  Prof. Guoming Chen for reminding us the diphoton excess observed by CMS collaboration, 
Prof. Xiaojun Bi, Yufeng Zhou, Pengfei Yin and Weihong Zhang for their instructions about DM indirect searches, 
and Dr. Liangliang Shang for his help on Monte Carlo simulation of sparticle searches. This work is
supported by the National Natural Science Foundation of China (NNSFC) under grant No. 11575053 and 11275245.

\end{document}